\documentclass[%
reprint,
nofootinbib,
amsmath,amssymb,
aps,
pra,
]{revtex4-2}

\usepackage{graphicx}
\usepackage{dcolumn}
\usepackage{bm}
\usepackage{hyperref}

\hypersetup{
  colorlinks   = true, 
  urlcolor     = blue, 
  linkcolor    = blue, 
  citecolor    = blue  
}

\usepackage{enumerate}
\usepackage{xcolor}

\usepackage{wasysym}

\newcommand{\comment}[1]{}

\newcommand{\eq}[1]{Eq.~\eqref{#1}}
\newcommand{\eqs}[1]{Eqs.~\eqref{#1}}
\newcommand{\ud}{\mathrm{d}}

\newcommand{\rg}{r_{g}}

\newcommand{\rp}{r_\mathrm{p}}
\newcommand{\rt}{r_{\rm t}}

\newcommand{\mbh}{M}
\newcommand{\mb}{m_{\rm b}}
\newcommand{\Eb}{E_{\rm b}}
\newcommand{\Lb}{L_{\rm b}}
\newcommand{\eb}{e_{\rm b}}
\newcommand{\Ec}{E_{\rm cm}}
\newcommand{\Lc}{L_{\rm cm}}
\newcommand{\xc}{\bm{x}_{\rm cm}}
\newcommand{\dxc}{\dot{\bm{x}}_{\rm cm}}

\newcommand{\rc}{r_{\rm cm}}

\def\Msun{M_{\odot} }

\begin{document}

\title{Relativistic tidal separation of binary stars by supermassive black holes}

\author{Luis A.~Manzaneda}
\affiliation{Facultad de Ciencias Puras y Naturales, Universidad Mayor de San Andr\'es,
Campus Universitario, Calle 27 Cota-Cota, La Paz, Bolivia}

\author{C\'esar O.~Navarrete}
\affiliation{Instituto de Ciencias Nucleares, Universidad Nacional Autónoma de M\'exico,
Circuito Exterior C.U., A.P.~70-543 Coyoac\'an, Ciudad de M\'exico D.F. 04510, M\'exico}

\author{Emilio Tejeda}
\thanks{Corresponding author: \href{mailto:emilio.tejeda@umich.mx}{emilio.tejeda@umich.mx} }
\affiliation{CONAHCYT, Avenida Insurgentes Sur 1582,
Cr\'edito Constructor, 03940 Ciudad de M\'exico, M\'exico }
\affiliation{Instituto de F\'isica y Matem\'aticas, Universidad Michoacana de San Nicol\'as
de Hidalgo, Edificio C-3, Ciudad Universitaria, 58040 Morelia, Michoac\'an, M\'exico}

\date{\today}

\begin{abstract}
A binary stellar system that ventures too close to a supermassive black hole can become tidally separated. In this article, we investigate the role of relativistic effects in these encounters through 3-body simulations. We use the Hybrid Relativistic-Newtonian Approximation (HRNA), which combines the exact relativistic acceleration from a Schwarzschild black hole with a Newtonian description of the binary's self-gravity. This method is compared against Newtonian and Post-Newtonian (1PN) simulations. Our findings show good agreement between HRNA and 1PN results, both of which exhibit substantial differences from Newtonian simulations. This discrepancy is particularly pronounced in retrograde encounters, where relativistic simulations predict up to $30\%$ more separation events and an earlier onset of binary separation ($\beta=2$ compared to $2.5$ in Newtonian simulations, with $\beta$ the impact parameter). Additionally, the HRNA model predicts about 15$\%$ more potential extreme mass ratio inspirals and generate a higher number of hypervelocity star candidates, with velocities up to 2,000 km/s faster than those predicted from Newtonian simulations. Furthermore, compared to Newtonian cases, relativistic encounters are more likely to result in direct stellar collisions and binary mergers.
\end{abstract}


\maketitle



\section{Introduction}
\label{S1}

When a binary stellar system ventures sufficiently close to a black hole (BH), the tidal forces exerted by the BH can become comparable to the binary's self-gravity, leading to tidal separation \cite{antonini10}. In scenarios where the binary approaches the BH along a parabolic orbit, this separation results in one member of the binary being gravitationally bound to the BH, while the other becomes unbound, receding with velocities reaching several thousand km/s. 

Stars ejected from such events are commonly referred to as hypervelocity stars (HVS), a concept first proposed by Hills in 1988 \cite{hills88}. 
The observational discovery of HVSs, initially reported by Brown \cite{brown05} and recently summarized in Ref.~\cite{brown15}, has sparked significant interest in the astrophysical community (e.g.~Refs.~\cite{yu03,bromley06,Zhang10,rossi14}). Alternative mechanisms for producing fast-moving, unbound stars from galactic centers have also been discussed in the literature (see, e.g.~Refs.~\cite{manukian13,guillochon_loeb,Rasskazov19}). 

Several studies have focused instead on the fate of the star that remains bound to the SMBH. For instance, in the context of elucidating the origin of stellar populations in close orbits around SMBHs, such as the S-stars around Sgr A* \cite{Zhang13,generozov20}. If this former member of the binary system is a compact object, it may evolve into a source of extreme mass-ratio inspirals (EMRI) \cite{miller05,addision19}, a phenomenon that could soon be observed with LISA \cite{AS23}. 

Tidal separation encounters can give rise to other types of interesting phenomena, including double tidal disruption events \cite{mandel15,bonnerot19}, stellar collisions \cite{Ginsburg07}, and binary mergers \cite{antonini12,bradnick17,FK19}. Comprehensive explorations of the relevant parameter space  for these encounters have been conducted through numerical simulations of the corresponding (full or restricted) 3-body problem (e.g.~Refs.~\cite{Zhang10,sari10,kobayashi12,rossi14,bkrs18,sari10,kobayashi12,addision19}). However, these studies have been limited to encounters occurring within a Newtonian regime.  

In this work, we consider tidal separation events occurring within the relativistic regime near a central non-rotating BH characterized by Schwarzschild spacetime.  We adopt the approximated relativistic treatment introduced by Tejeda et al.~in Ref.~\cite{tejeda17}, referred to hereafter as the Hybrid Relativistic-Newtonian Approximation (HRNA).  

This approach combines exact relativistic accelerations due to the central BH with a Newtonian description of the binary self-gravity. As demonstrated in Ref.~\cite{tejeda17}, HRNA accurately captures relativistic effects, particularly in scenarios where the gravitational field of the central BH dominates the curvature of the overall spacetime.  

To identify the primary features of a relativistic description of binary tidal separations, we perform a preliminary parameter space exploration. This entails comparing the outcomes of HRNA simulations with  their Newtonian counterparts. Furthermore, we validate the HRNA method by performing first-order Post-Newtonian (1PN) simulations for the same set of encounters. Our analysis reveals a robust agreement between HRNA and 1PN outcomes, highlighting consistent predictions from both relativistic approaches. Notably, these predictions significantly diverge from results obtained through Newtonian simulations of retrograde encounters. 

An advantage of the HRNA method over the PN formalism lies in its versatility, enabling its application to any black hole (BH) spacetime, whether prescribed analytically or numerically. While our focus in this study centers on tidal separations by a Schwarzschild BH, HRNA offers the flexibility to explore other spacetime metrics, such as those surrounding a rotating BH (Kerr spacetime), as demonstrated in Ref.~\cite{tejeda17,gafton19}. 

This versatility stands in contrast to the PN formalism, where the accuracy hinges on the order of the expansion utilized, with higher orders necessitating increasingly complex computations \cite{PW14}. 

The paper is organized as follows. In Section~\ref{S2} we describe the problem at hand, focusing on identifying the regime where relativistic effects become significant. Here, the HRNA method is introduced, detailing its principles and outlining self-consistency and validation tests to ensure its reliability. Section~\ref{S3} follows, detailing the numerical setup employed in the study. This includes a comprehensive description of the parameter space explored and its relevance to various astrophysical scenarios. In Section ~\ref{S4}, the simulation results are presented and analyzed. Key metrics such as separation fraction and properties of separated binaries are discussed, along with a thorough examination of different outcomes such as hypervelocity stars, possible EMRI sources, stellar collisions, and binary mergers. Finally, Section~\ref{S5} summarizes the findings and conclusions of the study. 

\section{Problem description}
\label{S2}

Consider a binary system comprising two stars with masses $m_1$ and $m_2$, with an initial separation $a_0$, approaching a BH of mass $\mbh$. Similar to tidal disruption events of individual stars, the relative strength of a given encounter can be characterized in terms of the impact parameter 
\begin{equation}
 \beta = \frac{\rt}{\rp},
 \label{e1.0}
\end{equation}
where $\rp$ is the distance of closest approach between the binary's center of mass (CM) and the BH (pericenter distance), and $\rt$ is the tidal radius defined as 
\begin{equation}
\begin{split}
\rt & \equiv \left(\frac{\mbh}{\mb}\right)^{1/3} a_0\\
& \simeq 101.3 \left(\frac{\mbh}{10^6\;\Msun}\right)^{-2/3} 
\left(\frac{\mb}{\Msun}\right)^{-1/3}
\left(\frac{a_0}{0.01\, \mathrm{au}}\right) \,\rg,
\end{split}
\label{e1.1}
\end{equation}
where $\mb = m_1 + m_2$  is the total binary mass and \mbox{$\rg\equiv G \mbh/c^2$} is the gravitational radius of the BH.\footnote{Unless otherwise stated, for the rest of this work  we adopt the geometrized unit system where $G=c=1$. Since then $\rg = \mbh$, we shall use these two variables interchangeably.} The tidal radius has been expressed in terms of physical parameters adequate for describing a binary system formed by two 0.5$\Msun$ white dwarfs.

In general, tidal binary separation is expected for encounters with $\beta\gtrsim 1$. However, it is important to stress that $\rt$ provides only an order-of-magnitude estimate of where tidal forces significantly influence binary dynamics. As previously discussed in the literature \cite{hut83}, a critical factor determining the outcome of such encounters is the spatial orientation with which the binary approaches the central object. 

\begin{figure}
 \begin{center}
\includegraphics[width=\linewidth]{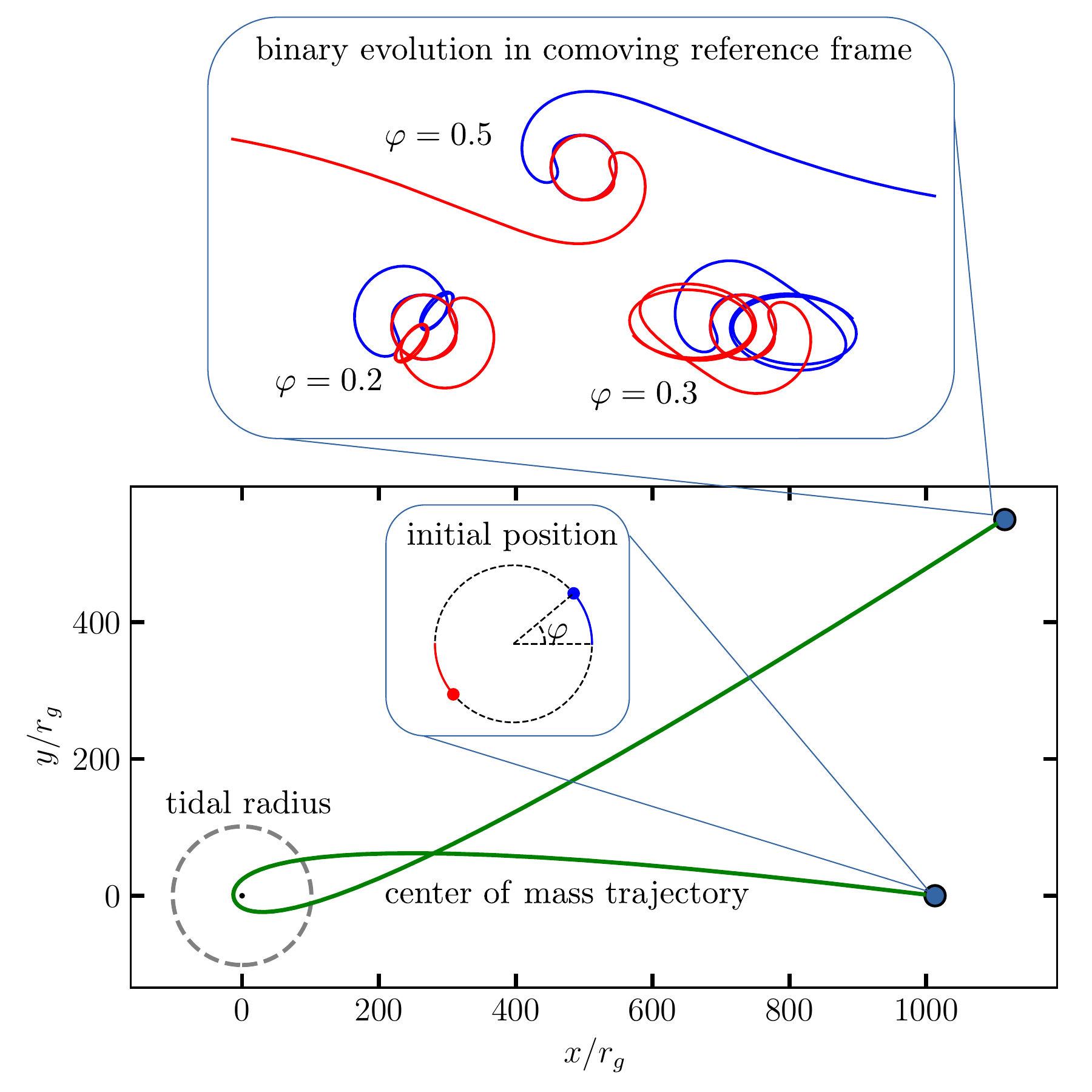}
 \end{center}
\caption{Examples of different possible outcomes of the encounter between a binary system and a supermassive BH. The central object is a Schwarzschild BH with $\mbh = 10^6\,\Msun$, with a corresponding gravitational radius of $\rg \simeq 0.01\,$au. The binary consists of two equal mass stars with $m_1 = m_2 = 0.5\,\Msun$, placed initially on a circular orbit separated by $a_0 = 0.01\,\mathrm{au} \simeq \rg$. The corresponding tidal radius is of $\rt = 101.3\,\rg$, while the impact parameter for this encounter is $\beta = 8$. The binary is on a retrograde orbit co-planar with the CM trajectory. The figure shows the trajectory of the binary's CM around the BH (the relativistic precession of pericenter is clearly visible). The magnified inset shows the binary system in its initial position (in this case at a distance $r_0 = 10\,\rt$). The top panel shows the whole binary evolution as seen from the CM reference frame. The outcome of three encounters with different initial phases ($\varphi$) are shown: $\varphi=0.5$ leading to  binary separation, $\varphi=0.3$ resulting in a surviving binary that has become wider, and $\varphi=0.2$ leaving behind a surviving binary that has become more compact. Given the very eccentric orbit of the latter, this encounter might actually result in a stellar collision once the finite radius of each star is considered. }
\label{fig1}
 \end{figure}

Consider, for example, the scenarios depicted in Fig.~\ref{fig1}, where we illustrate a binary's trajectory with $\beta=8$, along with the outcomes of three encounters viewed from a reference frame comoving with the binary's CM. All three encounters share the same parameters, except for the initial angular phase $\varphi$ of the binary. As evidenced by this figure, even minors variations in $\varphi$ can decisively impact the binary's fate. 

\begin{figure}
 \begin{center}
  \includegraphics[width=\linewidth]{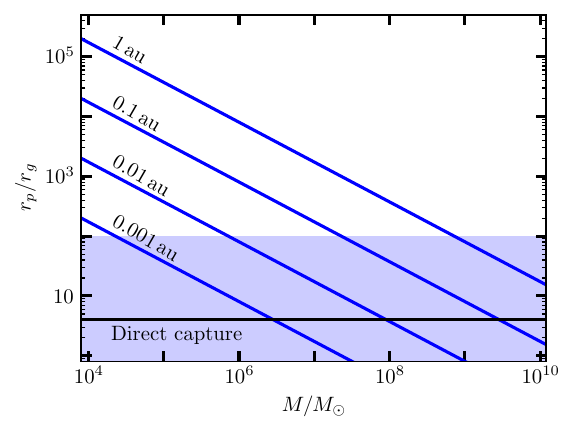}
 \end{center}
\caption{Parameter space for the tidal separation of a binary system, consisting of two equal $1\,\Msun$ stars with varying binary semi-major axes. The vertical axis represents the pericenter distance of the parabolic trajectory along 
which the binary approaches the central BH, whose mass $\mbh$ is shown on the 
horizontal axis. The blue lines indicate the tidal radii $\rt$ for 
different binary separations. The black horizontal line marks the limit of direct capture ($4\,\rg$) in Schwarzschild spacetime. General relativistic effects become relevant (order 1 percent or higher) for encounters that occur within the blue-shaded region.} 
\label{fig2}
 \end{figure}

In broad terms, general relativistic effects are expected to become significant for tidal separation events when the pericenter distance is less than $100\,\rg$ (see Ref.~\cite{tejeda17}, for a detailed description of the relativistic effects relevant for tidal encounters). In Fig.~\ref{fig2}, we display $\rt$ for various BH masses and initial binary separations. For instance, considering a supermassive BH with $\mbh = 10^6\;\Msun$, relativistic effects are predicted to be relevant for close binaries with initial semi-major axes $a_0 < 0.01$ au. Furthermore, the importance of these effects increases for a broader range of binary separations as we consider larger BH masses. 

General relativity naturally limits the maximum attainable impact parameter before the binary plunges whole onto the BH's event horizon. For parabolic-like trajectories (zero asymptotic velocity), this limit corresponds to the radius of the marginally bound circular orbit and is equal to 4\,$\rg$ for a Schwarzschild BH. From this condition we find
\begin{equation}
 \beta_{\rm max} \simeq 25.3 \left(\frac{\mbh}{10^6\;\Msun}\right)^{-2/3} 
\left(\frac{\mb}{\Msun}\right)^{-1/3}
\left(\frac{a_0}{0.01\, \mathrm{au}}\right).
\label{bmax}
\end{equation}

As evidenced by the strong pericenter shift of the CM trajectory in Fig.~\ref{fig1}, these encounters correspond to simulations run with the HRNA method that we describe in further detail next.

\subsection{Hybrid relativistic-Newtonian approximation}
\label{S2.2}

In the present work we treat the interaction between the binary stars as an external, Newtonian force acting on top of the fixed background metric of a Schwarzschild BH. In what follows, we refer to this treatment as Hybrid Relativistic-Newtonian Approximation (HRNA). A test particle, with rest mass $m$ and coordinates $x^\alpha$ in a curved spacetime subjected to an external force $f^\alpha$, obeys the equation of motion\footnote{Here and in what follows we adopt Einstein's convention of summation over repeated indices, with Greek indices running over spacetime components and Latin indices over spatial ones.} \cite{mtw}
\begin{equation}
\frac{\ud^2 x^\alpha}{\ud \tau^2} + \Gamma^\alpha_{\mu\nu} \frac{\ud x^\mu}{\ud \tau} \frac{\ud x^\nu}{\ud \tau} =\frac{1}{m}f^\alpha,
\label{eacc}
\end{equation}
where $\tau$ is the proper time of the test particle and $\Gamma^\alpha_{\mu\nu}$ are the Christoffel symbols associated to the spacetime metric $g_{\mu\nu}$. We shall thus take as external force
\begin{equation}
 f^\alpha = P^{\alpha\beta} \frac{\partial \Phi}{\partial x^\beta},
 \label{e.f}
\end{equation}
where 
\begin{equation}
    P^{\alpha\beta} = g^{\alpha\beta} + U^\alpha U^\beta 
\end{equation}
is the projection tensor that ensures that the resulting 4-acceleration $f^\alpha/m$ is always orthogonal to the 4-velocity $U^\alpha=\ud x^\alpha/\ud \tau$. On the other hand, in this work we model the binary's self-gravity via the Newtonian gravitational potential 
\begin{equation}
 \Phi = -\frac{ m_1\,m_2}{|\bm{x}_1 - \bm{x}_2|},
 \label{phi}
\end{equation}
where $\bm{x}$ refers to the spacial part of $x^\mu$ and $|\bm{x}_1 - \bm{x}_2|$ is the Euclidean distance between the two stars, that we treat here as point particles. 

Even though introducing the potential in \eq{phi} is in violation of general relativity's covariance principle and thus, constitutes at most an ad-hoc approximation, it is still accurate for situations in which the mass of the central BH is much larger than the mass-energy content of the surrounding matter. This has been shown in Refs.~\cite{tejeda17} where the same methodology was employed to study the tidal disruptions of individual stars by supermassive BHs.\footnote{Note that the method employed in Ref.~\cite{tejeda17} does not consider the projection operator, nevertheless it is simple to check that the resulting evolution equations (with respect to the coordinate time $t$) are the same in both cases.}

Given that the proper time of each individual star will progress at slightly different rates, we follow Ref.~\cite{tejeda17} and rewrite the spatial part of \eq{eacc} in terms of the coordinate time $t$ as
\begin{equation}
\begin{split}
\ddot{x}^i =& -\left(g^{i \lambda}-  g^{0\lambda}\,\dot{x}^i\right)\\
&\times
\bigg[ \frac{1}{m\Gamma^2}\frac{\partial \Phi}{\partial x^\lambda} + 
\left( \frac{\partial g_{\mu\lambda}}{\partial x^\nu} - 
\frac{1}{2}\frac{\partial g_{\mu\nu}}{\partial x^\lambda} \right)\dot{x}^\mu\, 
\dot{x}^\nu \bigg],
\label{CoordtEqs}
\end{split}
\end{equation}
where an over-dot indicates derivative with respect to $t$, $m$ refers to the mass of the star in question (i.e.~$m_1$ or $m_2$), $\Phi$ is the binary's self-gravitational field defined in \eq{phi}, and $\Gamma$ is the generalized Lorentz factor of the particle given by
\begin{equation}
\Gamma = \left(-g_{\mu\nu}\,\dot{x}^\mu\, \dot{x}^\nu\right)^{-1/2}.
\label{gamma}
\end{equation}

For numerically implementing this equation, we take 
\begin{equation}
 -\frac{1}{m}\frac{\partial\Phi}{\partial x^i} = a_i ,
\end{equation}
with $a_i$ the usual Newtonian acceleration due to the binary interaction. Also note that, according to \eq{phi},  $\partial \Phi/\partial t = 0$.

\subsection{Self-consistency and validation tests}
\label{S2.3}

A self-consistency criterion that was proposed in Ref.~\cite{tejeda17} for the validity of the HRNA method is to ask for the length scale on which self-gravity acts to be much smaller than the local radius of curvature $\mathcal{R}$ due to the BH. For a Schwarzschild BH we have \mbox{$\mathcal{R} = (r/\rg)^{3/2}\,\rg$} and, thus, we need to verify that for all of the encounters considered in this work the following expression is satisfied
\begin{equation}
 \alpha = \frac{a}{\mathcal{R}} \ll 1 , 
 \label{alpha}
\end{equation}
where, provided that the binary remains self-bound, $a$ is the binary separation. Once the binary is separated, the acceleration due to the self-gravity potential becomes negligible and each star effectively moves along independent geodesic trajectories of Schwarzschild spacetime. 

As we discuss in further detail in Appendix~\ref{Aalpha}, \mbox{$\alpha\sim10^{-5}$} for most of the simulation time of the encounters studied in this article. Nevertheless, for some of the deepest encounters, this parameter can reach values as large as $0.2$  during a short interval of time centered around the binary’s pericenter passage.

In Ref.~\cite{tejeda17} several tests were presented as proof of concept of the HRNA method. These tests included:
\begin{itemize}
    \item Geodesic motion limit. In both Schwarzschild and Kerr spacetimes, it was demonstrated that the exact geodesic motion is recovered under circumstances where self-gravity is negligible compared to the gravitational attraction due to the central BH. This typically applies to the CM motion when the system is well outside its tidal radius, or to the motion of individual particles after the system has been completely disrupted.   
    \item Comparison with previous studies of relativistic tidal disruption events (see, e.g.~Refs.~\cite{laguna93a,kobayashi04,kesden12,GT15}).
    \item Covariance principle. To evaluate the extent to which the HRNA method adheres to the covariance principle of general relativity, the outcome of several simulations with the same exact invariant initial conditions were compared using two different coordinate systems (Kerr-Schild and Boyer-Lindquist). 
\end{itemize}

As an additional validation test, using a harmonic oscillator as a toy model of a self-interacting system, in Appendix~\ref{AH} we demonstrate that the HRNA method correctly captures the relativistic dilation of time. Specifically, we show that the time scale associated to the internal interaction of a bound system dilates in the same manner as that of a free-falling, virtual test particle positioned at the system's CM.\footnote{This observation is made from the perspective of a physical observer located asymptotically far from the central BH.} This confirms that the HRNA method does not introduce any artificial amplification of tidal effects attributable to relativistic time dilation.

\section{Numerical setup}
\label{S3}

As previously mentioned, this work involves numerical studies of close encounters between a binary stellar system, modeled as two point masses, and a supermassive BH. In order to explore the role of general relativistic effects in this kind of interactions, we compare the outcome of each encounter when evolved under three different gravity descriptions:
\begin{enumerate}[i)]
    \item Newton's gravity law.
    \item The hybrid relativistic-Newtonian approximation (HRNA) described in Section \ref{S2}.
    \item First-order, post-Newtonian interactions (1PN) \cite{Straumann2013,Merritt2016}. A summary of the 1PN evolution equations for a restricted three-body system is given in Appendix~\ref{AA}.
\end{enumerate}

The gravitational encounter between three bodies comprises at least 21 different parameters (mass, initial position and velocity for each particle). In this work we  reduce the number of degrees of freedom by adopting the following restrictions:
\begin{enumerate}[i)]
    \item The central BH is considered to be much more massive than the binary system. Therefore, we neglect any kind of back-reaction from the binary on to the central BH and fix its position at the origin of a global reference frame.

    \item The binary system consists of two-equal mass stars ($m_1=m_2$) in a circular orbit around each other in the beginning of the simulation. We call $a_0$ their initial separation and $\varphi$ the initial orbital phase (true anomaly). The angle $\varphi$ is measured with respect to the $x$ axis of the global reference frame.    
    
    \item The binary system approaches the central BH following a parabolic-like trajectory, that is, the binary's CM has a zero kinetic energy asymptotically far away from the central object.

    \item The binary and CM trajectories are co-planar, i.e.~all the encounters are restricted to one plane that, for simplicity, we take as $z=0$.
\end{enumerate}

The previous restrictions leave us with 7 model parameters: BH mass $\mbh$, total binary mass $\mb$, initial binary separation $a_0$, impact parameter $\beta$, initial orbital phase $\varphi$, initial binary distance from the BH $r_0$, and sense of rotation of the binary with respect to its CM motion (i.e. prograde or retrograde motion). Further details for implementing the initial conditions for each gravity law are given in Appendix~\ref{AB}.

Since the binary is assumed to approach the central BH from infinity, our results should in principle be independent from $r_0$ provided that $r_0 \gg \rt$.\footnote{As we define the initial orbital phase $\varphi$ with respect to the $x$ axis of the global reference frame, the outcome of an individual encounter certainly depends on the initial distance $r_0$ from which the binary is released. However, we are not concerned with the outcome of any given encounter with a particular initial $\varphi$, but rather, with the outcome of a large ensemble of encounters spanning all possible initial phases $\varphi\in[0,\,2\pi]$.}  We take $r_0 = 50\,\rt$ as fiducial value for the initial distance of the CM trajectory. From this point, we evolve the system in time as the binary approaches the central BH, passes through pericenter, and then recedes until the distance from the binary’s CM to the origin exceeds $150\,\rt$, at which point we terminate the simulation. An analysis concerning the convergence of the simulations with respect to the initial and final points is discussed in Appendix~\ref{AC}.

The parameter space can be further reduced to five dimensions by noticing that this problem can be rendered scale free by adopting $\mbh$ as unit of mass and length, i.e.~for a given sense of rotation and fixed values of $\beta$ and $\varphi$, encounters with same mass ratios $\mb/\mbh$ and same initial distances ($a_0/\rg$ and $r_0/\rg$) are equivalent to one another.

We will consider as fixed model parameters \mbox{$\mb = 10^{-6} \mbh$} and  $a_0 = 1.013\,\rg$, and study the resulting three-dimensional parameter space: $\beta$ and $\varphi$ as continuous variables, and sense of rotation as a discrete parameter corresponding to either prograde or retrograde orbits.
From \eqs{e1.1} and \eqref{bmax}, with these parameters we have a tidal radius of $\rt = 101.3\,\rg$ and a maximum impact parameter of $\beta_{\rm max} = 25.3$. However, as shown in Appendix~\ref{Aalpha}, the self-consistency parameter $\alpha$ in \eq{alpha} is an increasing  function of $\beta$ and, already for $\beta=10$, a substantial fraction of encounters reach $\alpha \gtrsim 0.1 $. For this reason, we restrict our exploration to $\beta \leq 10$.

With these fixed parameters, the initial (unperturbed) binary period is given by
\begin{equation}
\begin{split}
    T_0 & = 2\pi\sqrt{\frac{a_0^3}{m_b}} \\
    & = 6,406\, M \simeq 8.8\, \left(\frac{\mbh}{10^6\;\Msun}\right) \, \mathrm{hr},
\end{split}
\end{equation}
while an estimate of the time taken by the binary to reach $r_p$ from $r_0$ is given by
\begin{equation} 
\begin{split} 
\Delta t & \simeq \frac{4}{3}\left(\frac{r_0}{2}\right)^{3/2} M \\ 
& = 170,000\, M \simeq 232 \left(\frac{M}{10^6\Msun}\right) \mathrm{hr}.
\end{split} 
\end{equation}

By adopting re-scaled physical units, our fixed model parameters are suitable for describing a continuum of possible encounters. From this broad spectrum, we can highlight the following representative types of encounters:
\begin{itemize}
 \item[] {\bf type 1:} $\mbh = 10^6\Msun$, $\mb = 1\,\Msun$, $a_0 = 0.01\,$au,

 \item[] {\bf type 2:} $\mbh = 10^7\Msun$, $\mb = 10\,\Msun$, $a_0 = 0.1\,$au,

 \item[] {\bf type 3:} $\mbh = 10^8\Msun$, $\mb = 100\,\Msun$, $a_0 = 1.0\,$au,
\end{itemize}
with type 1 consisting of a close binary composed of two $0.5\,\Msun$ white dwarfs; type 2 consisting of two main sequence $5\,\Msun$ stars; and type 3 consisting of two $50\,\Msun$ BHs (with an initial separation wide enough to ensure that their orbital motion can be treated within the Newtonian regime).

\section{Simulation results}
\label{S4}

We explore a broad parameter space by sampling 200 equally spaced values of the impact parameter $\beta$, ranging from 0.1 to 10. For each value of $\beta$, we consider 400 initial phases $\varphi$, uniformly distributed between 0 and $\pi.$\footnote{We only need to consider half of the full interval of the initial phase $\varphi\in[0,2\pi]$ given that the assumption $m_1 = m_2$ ensures its symmetry.} For every pair of parameters $(\beta,\varphi)$, we run six independent simulations, one for each gravity law (Newtonian, HRNA and 1PN) and for both prograde and retrograde encounters. This approach resulted in a total of 480,000 si\-mu\-lations performed to conduct this initial exploration. 

The simulations are carried out with the N-body code {\sc rebound} \citep{rein12}, employing the 15th-order, implicit integrator with adaptive time stepping IAS15 developed by \citet{rein15}.\footnote{{\sc rebound} is an open-source code that can be downloaded freely at \url{http://github.com/hannorein/rebound}.} We have augmented the code by introducing a modulus that computes the acceleration due to a central BH according to both HRNA and 1PN descriptions. We have compared the outcome of the simulations against an independent Julia code using the Tsit5 integrator \cite{Tsit5}, finding a close agreement between the two sets of results.

At every point along the trajectory of a given simulation, we store the positions $\bm{x}_{1},\ \bm{x}_{2}$ and velocities $\dot{\bm{x}}_{1},\ \dot{\bm{x}}_{2}$ of each particle. We then introduce the binary's CM and relative position vectors (as computed in flat spacetime):
\begin{gather}    
\xc = \frac{m_1\,\bm{x}_{1}+m_2\,\bm{x}_{2}}{m_1 + m_2} = \frac{\bm{x}_{1}+\bm{x}_{2}}{2},\\ 
\bm{x}_{12} = \bm{x}_{1}-\bm{x}_{2},
\end{gather}
and the associated velocities
\begin{gather}    
\dxc = \frac{\dot{\bm{x}}_{1}+\dot{\bm{x}}_{2}}{2},\\
\dot{\bm{x}}_{12} = \dot{\bm{x}}_{1}-\dot{\bm{x}}_{2}.
\end{gather}

From these quantities, we compute instantaneous values for the specific energy $\Eb$, angular momentum $\Lb$ and eccentricity $\eb$ of the binary system according to the Newtonian expressions
\begin{gather}
    \Eb = \frac{1}{2}\left|\dot{\bm{x}}_{12}\right|^2 - \frac{\mb}{\left|\bm{x}_{12}\right|},\\
    \Lb = \left|\bm{x}_{12}\times\dot{\bm{x}}_{12}\right|,\\
    \eb = \sqrt{1+\frac{2\Eb\Lb^2}{m_b^2}}.
    \label{ecc_b}
\end{gather}

On the other hand, employing appropriate expressions for each gravity law,\footnote{In the Newtonian case we employ the usual expressions for the motion of a test particle; for the relativistic approach we employ the exact expressions for conserved quantities in a Schwarzschild spacetime (see, e.g.~Ref.~\cite{tejeda17}); for 1PN interactions we employ \eqs{EIH1} and \eqref{EIH2}.} we also compute instantaneous values for the specific energies $E_1,\,E_2,\,\Ec$ and angular momenta $L_1,\,L_2,\,\Lc$ of each binary member as well as of a virtual test particle located at the CM, all of these measured with respect to the central BH. 

It is clear that neither the energies nor the angular momenta defined above are expected to be strictly conserved quantities, as the interactions between the binary and tidal forces continuously alter their values. However, at distances significantly greater than the tidal radius, the impact of tidal forces diminishes to the extent that the binary’s orbital motion can be considered effectively decoupled from the motion of the CM. As a result, we can expect that $\Eb$, $\Lb$, $\Ec$ and $\Lc$ will remain approximately constant for $\rc=|\xc|\gg\rt$. 

Note that this assumption should remain valid both as the binary system approaches the black hole at the start of the simulation and also at its conclusion, provided the binary survives the encounter as a bound system. However, the values of these quantities may differ between these two points.

We say that a binary has survived an encounter if, by the end simulation, $\Eb$ settles to a constant value $\Eb<0$ (in which case, according to \eq{ecc_b} $\eb<1$). Conversely, if $\Eb>0$ by the simulation's conclusion, the binary is considered to have separated, with each star now following independent trajectories as test particles about the central BH. Therefore, for separation encounters, we can expect $E_1$, $E_2$, $L_1$, $L_2$ to reach approximately constant quantities by the end of the simulation.

\begin{figure*}
  \begin{center}
   \includegraphics[width=0.9\textwidth]{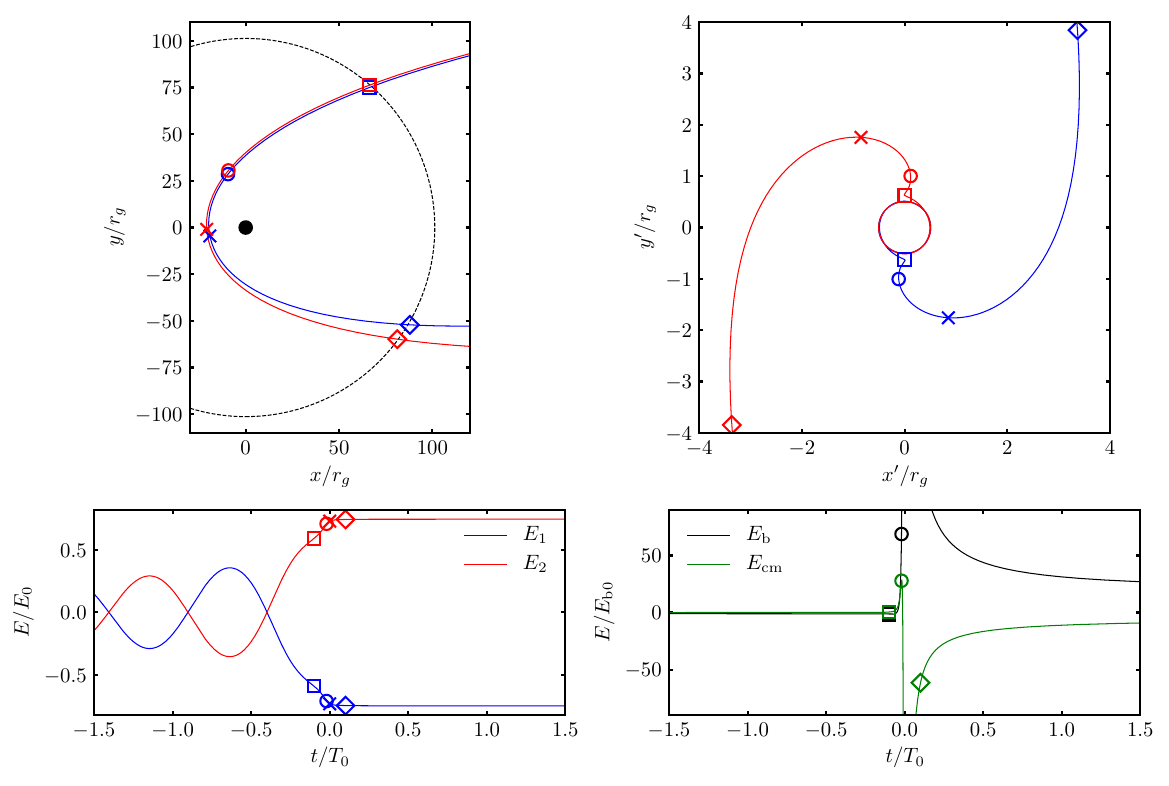}
  \end{center}
 \caption{ Example of a prograde separation encounter, corresponding to an HRNA simulation with $\beta = 5$ and $\varphi = 1.41$. {\bf Top-left panel}: binary trajectory as seen in the central BH's reference frame. {\bf Top-right panel}: binary trajectory as seen in the comoving reference frame. {\bf Bottom-left panel}: time evolution of the energies $E_1$ and $E_2$ expressed in units of $E_0 = \sqrt(Mm_b/(a_0\,\rt)) \simeq 10^{-4} $. {\bf Bottom-right panel}: time evolution of the energies $\Eb$ and $\Ec$ expressed in units of $E_{\mathrm{b}0} = m_b/(2\,a_0) \simeq 5\times10^{-7} $. Equal symbols on each panel correspond to the same instants of time. Time is measured in units of the binary's orbital period $T_0 = 6,406 M$. }
 \label{sep-P}
  \end{figure*}

An example of a separation encounter is illustrated in Fig.~\ref{sep-P}, depicting an HRNA simulation of a prograde encounter with $\beta=5$ and $\varphi = 1.41$. The top panels show the binary's trajectory as observed from the BH's reference frame (left-hand side) and from a comoving reference frame (right-hand side). The time evolution of the energies $E_1$ and $E_2$ is shown on the bottom left panel, while the bottom right panel shows the evolution of $\Eb$ and $\Ec$. 
From this figure we see that, for a separation encounter, $E_1$ and $E_2$ reach constant values (one positive and the other negative) at late times after separation.

Equal symbols on each panel correspond to the particles' positions at the following instants of time:
\begin{itemize}
    \item[$\square$] The binary enters the tidal radius.
    \item[$\times$] The binary reaches pericenter.
    \item[$\Diamond$] The binary exits the tidal radius.
    \item[$\fullmoon$] The binary separation exceeds $2 a_0$.
\end{itemize}

We take the last point as a proxy for indicating the moment of separation, defining the time $t_s$ such that
\begin{equation}
    a(t_s) = 2a_0.
    \label{ts}
\end{equation}
We have experimented with various multiplicative factors for determining this moment, yielding qualitatively similar results across the board. For the encounter represented in Fig.~\ref{sep-P}, we have
$$  t_s = -0.01\,T_0,\qquad  \rc(t_s) = 31\,\rg, $$  
where $t=0$ corresponds to the pericenter passage.
 
\begin{figure*}
  \begin{center}
   \includegraphics[width=0.9\textwidth]{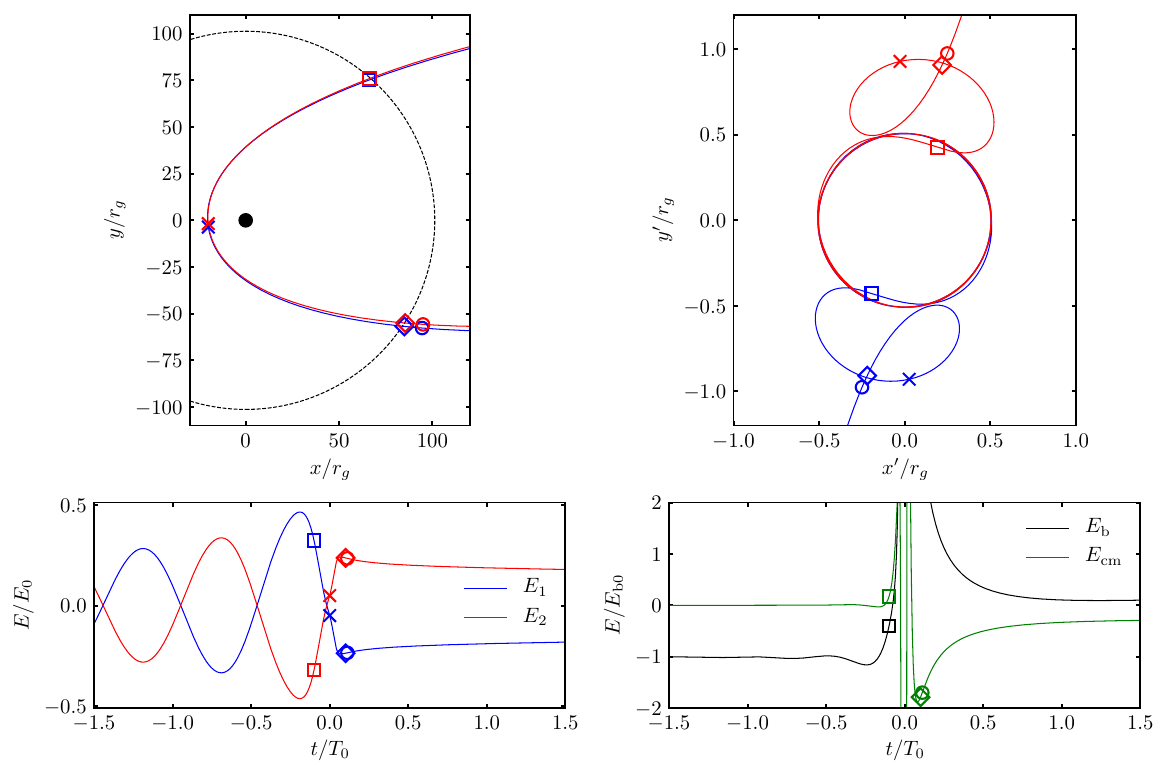}
  \end{center}
 \caption{ Example of a separation retrograde encounter, corresponding to an HRNA simulation with $\beta = 5$ and $\varphi = 1.57$.  Panels and units are the same as in Fig.~\ref{sep-P} . }
 \label{sep-R}
  \end{figure*}

In Fig.~\ref{sep-R}, we present another example of a separation encounter, this time along a retrograde trajectory with $\varphi=1.57$. This also corresponds to an HRNA simulation with $\beta=5$. In this instance, we have
$$  t_s = 0.08\,T_0,\qquad  \rc(t_s) = 107\,\rg, $$
indicating that the binary separated after pericenter, and at a greater distance from the BH compared to the previous example. 
From this figure, we can also notice that the binary's sense of rotation has reversed from retrograde to prograde before separation. This reversal in rotation is consistently observed in other separation encounters involving retrograde trajectories.

Finally, in Fig.~\ref{sep-P} we show an example of a retrograde surviving encounter with $\varphi=2.04$. As before this corresponds to an HRNA simulation with $\beta=5$. From this figure we can see that, in a surviving encounter, the energies $E_1$ and $E_2$ are constantly interchanging values in a symmetric way as the binary members orbit each other, with a marked shift in the binary period after pericenter passage at $t=0$. On the other hand, $\Eb$ is approximately constant at early and late times.
Also notice that, as in the previous example, the sense of rotation of the binary has reversed by the end of the interaction.

 \begin{figure*}
  \begin{center}
   \includegraphics[width=0.9\textwidth]{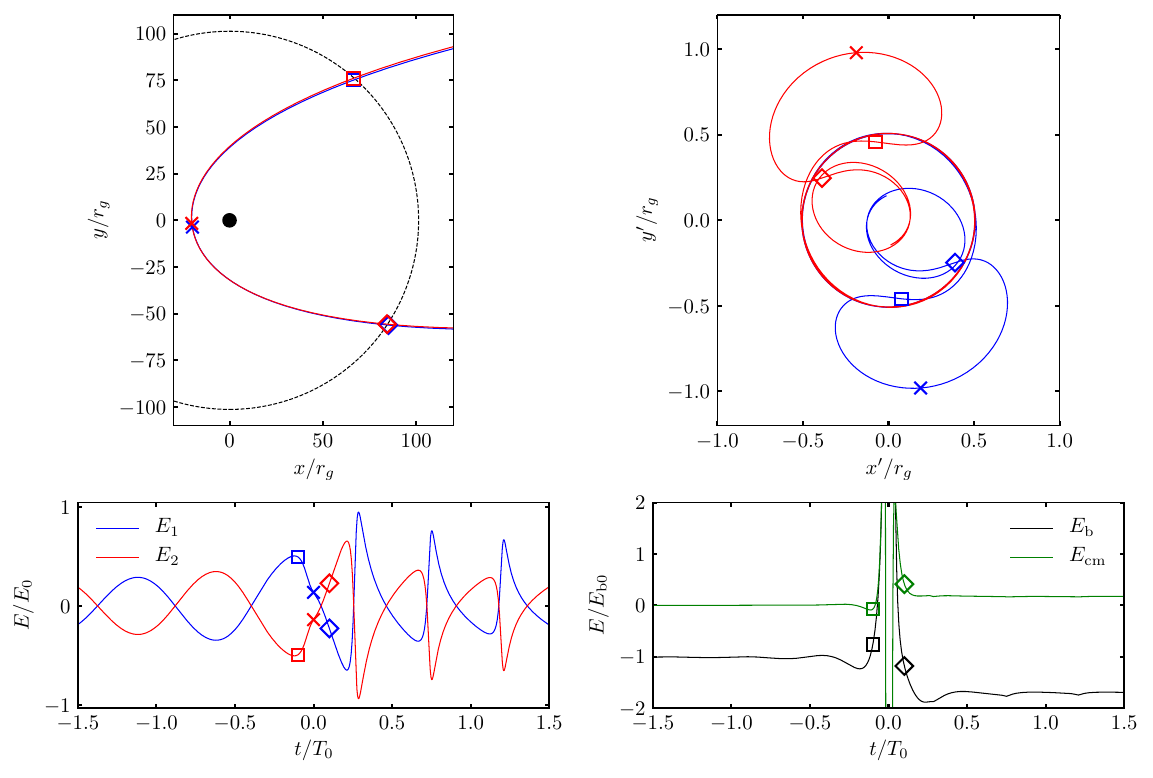}
  \end{center}
 \caption{ Example of a surviving retrograde encounter, corresponding to an HRNA simulation with $\beta = 5$ and $\varphi = 2.04$.  Panels and units are the same as in Fig.~\ref{sep-P}. }
 \label{sur-R}
  \end{figure*}%

\begin{figure}
 \begin{center}
\includegraphics[width=0.49\textwidth]{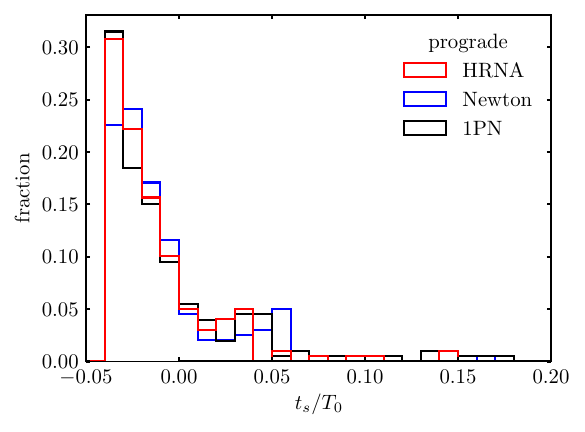}
\includegraphics[width=0.49\textwidth]{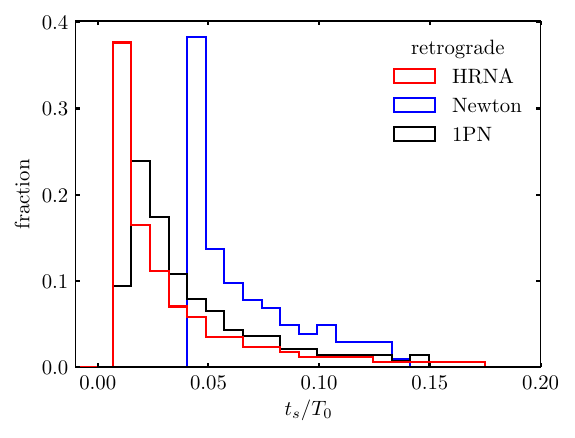} 
 \end{center}
\caption{Distribution of separation times $t_s$ for encounters with an impact parameter of $\beta = 5$ across the three gravity descriptions. Prograde encounters are shown in the top panel, while retrograde encounters are displayed in the bottom panel.} 
\label{fts}
 \end{figure}

In Fig.~\ref{fts}, we present a histogram depicting the separation times $t_s$ as defined in \eq{ts} for encounters with $\beta = 5$ across the three gravitational models. The fraction is calculated relative to the number of separation encounters in each case. The histogram for prograde encounters is shown in the top panel, while retrograde encounters are displayed in the bottom panel. It is notable that the majority of prograde encounters experience separation close to and before pericenter. Conversely, retrograde encounters occur later and after pericenter. Moreover, while the distributions for prograde encounters are similar across all three gravity laws, retrograde encounters under Newtonian physics show significant deviation from those predicted by HRNA and 1PN models. 

\subsection{Separation fraction}

With our exploration of 400 initial phases $\varphi$ for each impact parameter value $\beta$, we find that separation encounters cluster together within intervals of the form $[\varphi_i,\,\varphi_j]\in[0,\,\pi].$\footnote{For some values of $\beta$, separation encounters can be split into two intervals \mbox{$[0,\,\varphi_i]$} and \mbox{$[\varphi_j,\,\pi]$}. This still corresponds to a simply connected interval \mbox{$[\varphi_j,\,\pi+\varphi_i]$}, as the whole range of the initial phase is $\varphi\in[0,\,2\pi]$.}
We refine the search for the exact boundary where the transition from separation to surviving takes place by conducting a bisection root finding algorithm until two solutions $\varphi_1 \in[ \varphi_{i-1},\,\varphi_i]$ and $\varphi_2 \in[ \varphi_{j},\,\varphi_{j+1}]$ are found within a precision of $10^{-4}$. In average, each root is found after 7 iterations of the bisection algorithm. 

Once we identify the interval $[\varphi_1, \varphi_{2}]$ enclosing all of the separation encounters for a given $\beta$, we define the separation fraction as\footnote{In the case of prograde encounters, there are some values of $\beta$ with two or more disjoint intervals enclosing separation encounters. In these cases, we take the length of the union of all these intervals as numerator of \eq{sf}. }
\begin{equation}
    \mathrm{s.\,f.} = \frac{\varphi_2 - \varphi_1 }{\pi}.
    \label{sf}
\end{equation}
As discussed in more detail in Appendix~\ref{AC}, the relative error of the separation fraction with respect to the initial distance $r_0$ corresponds to $\sim 10^{-3}$ for Newtonian and 1PN simulations and to $\sim 10^{-2}$ for HRNA simulations.

\begin{figure}
  \begin{center}
   \includegraphics[width=0.5\textwidth]{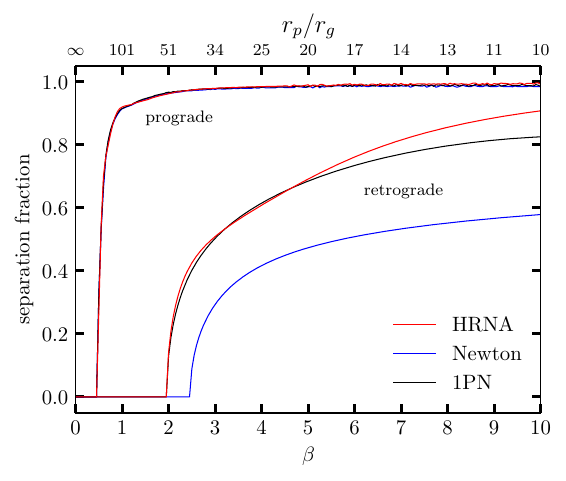}
  \end{center}
 \caption{ Separation fraction as a function of the impact parameter $\beta$ for prograde and retrograde orbits. Results are shown for each of the three gravity laws considered in this work. }
 \label{split}
  \end{figure}

In Fig.~\ref{split} we compare the resulting separation fraction as a function of the impact parameter for each gravity law, showing both prograde and retrograde orbits. From this figure we can see that prograde orbits are separated at smaller values of the impact parameter than those needed for retrograde orbits. This result is consistent with the findings of previous works \cite{sari10,addision19} and is primarily due to the fact that the binary rotation in prograde orbits facilitates the separation process, whereas in retrograde orbits it acts in opposition.

Notably, from Fig.~\ref{split} we see that the separation fraction of prograde encounters is essentially the same for all three gravity laws. In other words, most of the prograde encounters are being tidally separated in such a way that relativistic effects are hardly noticeable. 

On the other hand, relativistic effects are more prominent among retrograde encounters, as evidenced in Fig.~\ref{split} where we see a clear distinction between Newtonian, HRNA and 1PN encounters. For instance, we see that relativistic encounters (both HRNA and 1PN) start to break up at smaller impact parameters ($\beta\simeq2$) than Newtonian encounters ($\beta\simeq2.5$). Moreover, when considering a particular $\beta$, relativistic models produce separation fractions that exceed those from Newtonian models by up to 30$\%$.

From Fig.~\ref{split} we can also notice that HRNA and 1PN results are in good agreement with each other for values of $\beta\lesssim 5$. These two curves start to deviate from one another for the deepest encounters ($r_p\lesssim 20\,\rg$), a regime in which a 1PN description is insufficient to capture higher order relativistic effects.

\subsection{Properties of the separated binaries}

After a separation encounter, we can analyze in detail the kinematics of the former binary members, as each star is now moving along free-fall trajectories (geodesic motion). From this analysis we calculate conserved quantities such as energy and angular momentum, their associated radial turning points $r_a$ and $r_p$, and define the eccentricity of the resulting orbit as\footnote{Note that for bound trajectories $0<r_p<r_a$ and we can identify $r_p$, $r_a$ with pericenter and apocenter radii, respectively. Instead, for unbound trajectories $r_a<0$. In the relativistic case there is a third real root \mbox{$0<r_b<r_p$} of the radial motion, which is not relevant for our present analysis. }
\begin{equation}
    e = \frac{r_a - r_p}{r_a + r_p}.
\end{equation}

\begin{figure}
 \begin{center}
  \includegraphics[width=0.49\textwidth]{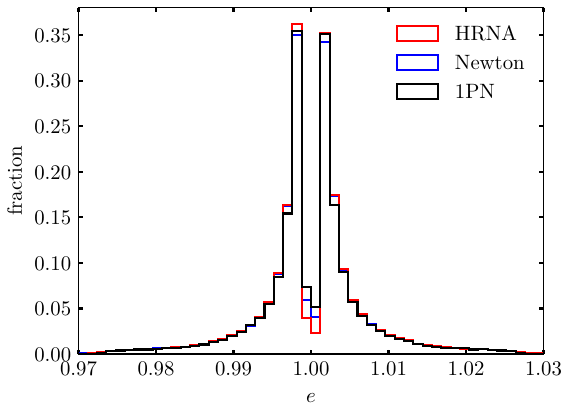}
  \includegraphics[width=0.49\textwidth]{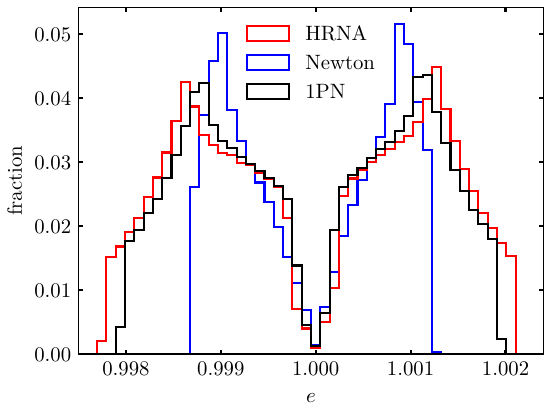}
  \caption{Final eccentricity distributions of the bound and ejected stars in separation encounters combining all impact parameters $\beta\in[0.1,\,10]$. For comparison, results are shown for Newtonian, HRNA and 1PN simulations. The top-panel shows prograde encounters and the bottom-panel shows retrograde ones. The fraction is normalized with respect to the total number of encounters (both surviving and separation).}
  \label{ecc12}
 \end{center}
\end{figure}

In Fig.~\ref{ecc12} we show the resulting eccentricity distributions for all values of $\beta\in[0.1,\,10]$. Prograde encounters are shown on the top-panel and retrograde ones on the bottom-panel. Values of $e<1$ correspond to bound trajectories, while $e>1$ to unbound ones. As for the separation fraction, from this figure we see that relativistic effects are more prominent among retrograde encounters, where we see a broader distribution in final eccentricities obtained from HRNA simulations as compared to Newtonian results. On the other hand, results from HRNA and 1PN simulations are in good agreement with each other.

The spread in eccentricities of former binary members observed in Fig.~\ref{ecc12} is consistent with the analytic estimate given by Ref.~\cite{addision19}:
\begin{equation}
    e \simeq 1 \pm 0.02\,\beta \left(\frac{\mb}{\Msun}\right)^{1/3} \left(\frac{\mbh}{10^6\Msun}\right)^{-1/3} .
\end{equation}

\begin{figure}
 \begin{center}
  \includegraphics[width=0.49\textwidth]{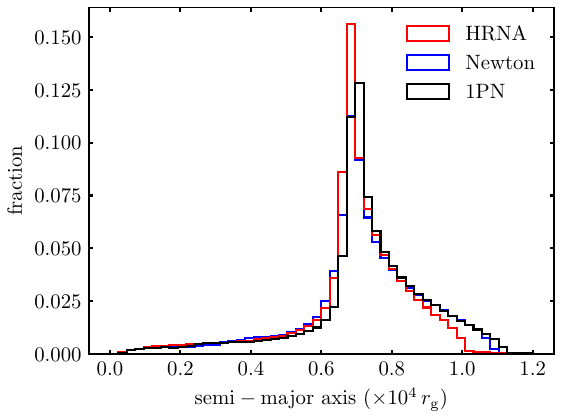}
  \includegraphics[width=0.49\textwidth]{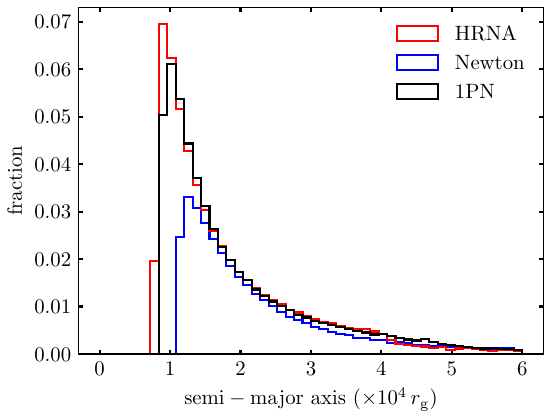}
  \caption{Histograms of the semi-major axis of the bound star in separation encounters. The top-panel shows prograde encounters and the bottom-panel shows retrograde ones. In both cases, the fraction is normalized with respect to the total number of encounters.}
  \label{semi-axis}
 \end{center}
\end{figure}

In Fig.~\ref{semi-axis} we show the distribution of the resulting semi-major axis $(r_a + r_p)/2$ of the former binary member that remains bound to the central BH after a separation encounter. In this case, we observe that relativistic simulations (both HRNA and 1PN) of retrograde encounters tend to produce bound stars with a smaller semi-major axis than Newtonian encounters. The properties of this population of bound stars are of interest in connection to the demographics of S-stars at the galactic center \cite{generozov20}.

\begin{figure*}
 \begin{center}
   \includegraphics[width=0.48\linewidth]{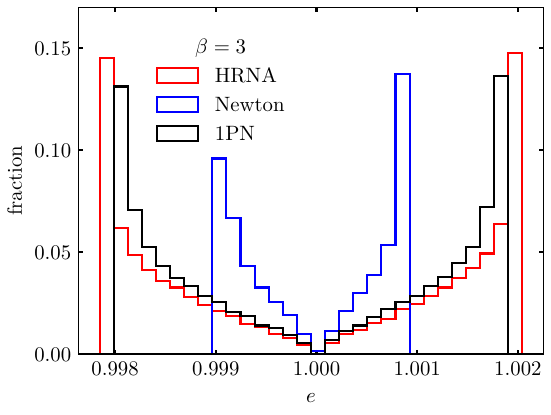}
  \includegraphics[width=0.48\linewidth]{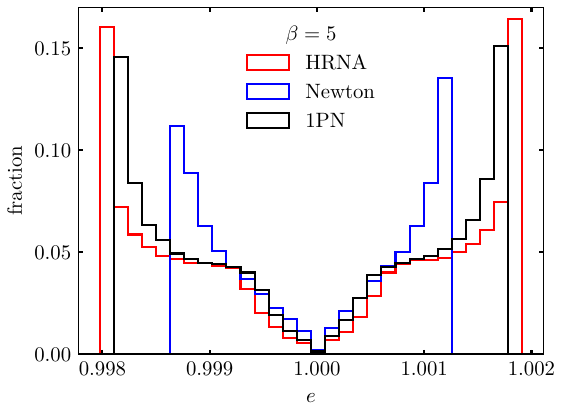}\\
  \includegraphics[width=0.48\linewidth]{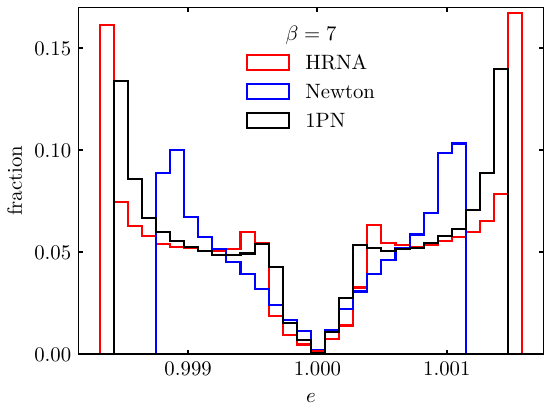}
  \includegraphics[width=0.48\linewidth]{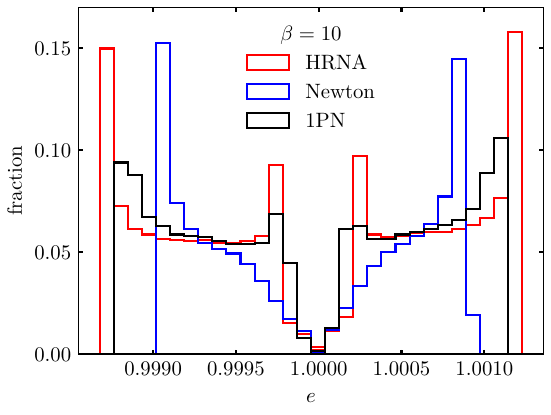}
 \end{center}
\caption{Histograms of eccentricities of separated binaries for Newtonian, HRNA and 1PN encounters in retrograde trajectories. Each panel corresponds to a different impact parameter $\beta$ as indicated on the legend title. The fraction is taken with respect to the total number of encounters. }
\label{eccs}
 \end{figure*}

Finally, in Fig.~\ref{eccs} we compare in detail the eccentricity distributions of retrograde trajectories for specific values of $\beta = 3,\, 5,\, 7,\, 10$. From this figure we notice that HRNA and 1PN simulations are in reasonable agreement with each other (especially for $\beta\leq 5$) while, as noted before, Newtonian simulations give place to a narrower eccentricity distribution. 


\subsection{Bound star as an EMRI source}

Under appropriate conditions, the former binary member that remains bound to the SMBH can become a source of gravitational waves as an EMRI. For the associated signal to lay within the observational window of the future LISA observatory, the bound star would need to be a compact object (either a neutron star or a solar mass BH), on a highly eccentric orbit \cite{AS23}. For instance, if the bound object is a 50$\Msun$ BH (type 3 encounter), the system is expected to merge due to gravitational wave emission within a Peters lifetime \cite{peters64} 
\begin{equation}
\begin{split}
    T_\mathrm{gw} & = \frac{3\,a_i^4}{85\, M^2 m}f(e_i) \\
    & \simeq 1.3 \,\frac{f(e_i)}{f(0.99)} 
    \left(\frac{a_i}{10^3\,\mathrm{au}}\right)^4\\
    & \quad \times 
    \left(\frac{M}{10^8\Msun}\right)^{-2}
    \left(\frac{m}{50\,\Msun}\right)^{-1}\,\mathrm{Myr},
\end{split}
\label{temri}
\end{equation}
where $m$ is the mass of the bound compact object, $e_i$ and $a_i$ are its (post-encounter) eccentricity and semi-mayor axis, and
\begin{equation}
    f(e) = (1-e^2)^{7/2}.
\end{equation}

\begin{figure}
  \begin{center}
   \includegraphics[width=0.49\textwidth]{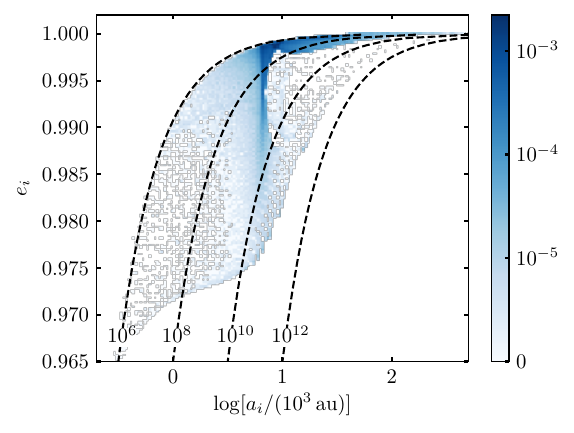}
   \caption{Distribution of compact object left bound to the SMBH (potential EMRI sources) as a function of the orbital parameters $(a_i,\,e_i)$ that result from both prograde and retrograde HRNA simulations. The color scale corresponds to the fraction with respect to the total number of encounters. The dashed, black lines indicate level sets  of equal Peters lifetime $T_\mathrm{gw}$ as indicated by each label in units of years. }
   \label{f_ae}
  \end{center}
 \end{figure}

In Fig.~\ref{f_ae}, we display the final distribution of $e_i$ and $a_i$ as derived from the HRNA simulations. Additionally, we include contour level lines representing equal $T_\mathrm{gw}$. Taking into account the distinct set of parameters used in our study, this figure can be compared to Fig.~6 of Ref.~\cite{sari10}. 
The visible gap in the distribution observed in our results arises because our analysis is confined to co-planar encounters, whereas Ref.~\cite{sari10} incorporates encounters with arbitrary orientations. 

\begin{figure}
  \begin{center}
   \includegraphics[width=0.49\textwidth]{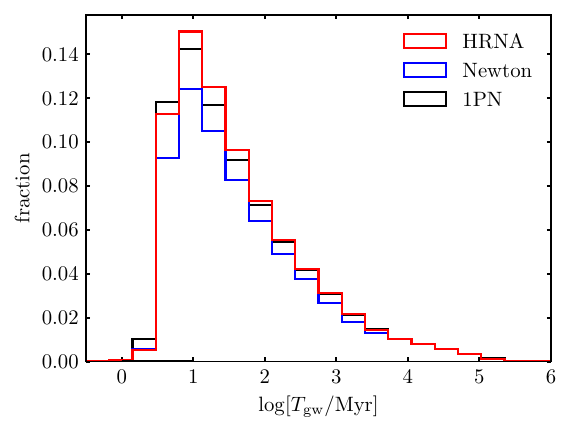}
   \caption{Distribution of $T_\mathrm{gw}$ for the potential EMRI sources derived from Newtonian, HRNA and 1PN simulations. For each model, the fraction is calculated relative to the total number of encounters. The results encompass all values of $\beta$ and include both prograde and retrograde trajectories.}
   \label{hist_EMRI}
  \end{center}
 \end{figure}

In Fig.~\ref{hist_EMRI}, we present the distribution of $T_\mathrm{gw}$,  expressed as a fraction of the total number of encounters, comparing the outcomes from Newtonian, HRNA and 1PN simulations across all values of $\beta$ and for both prograde and retrograde trajectories. The figure illustrates that while results from HRNA and 1PN  are in reasonable agreement, HRNA simulations predict approximately $15\%$ more EMRI candidates than the Newtonian simulations for Peters lifetimes ranging from $T_\mathrm{gw} = 10^6$ to $10^8\,$yr. 

\subsection{Velocity of the ejected star}

Given that the binary system approaches the BH along a parabolic-like trajectory ($\Ec = 0$ initially), the outcome of separation encounters invariably results in one former member of the binary remaining gravitationally bound to the BH (say, $E_1 < 0$), while the other is ejected out on a hyperbolic-like trajectory ($E_2 > 0$), in such a way that $E_1 + E_2 \simeq 0$. This symmetry between $E_1$ and $E_2$ is also apparent in their associated eccentricity values, as illustrated above in Fig.~\ref{eccs}.  

We define the asymptotic velocity $v_\infty$ of the ejected star, an HVS candidate, according to
\begin{equation}
    v_\infty = \sqrt{2\max(E_1,\,E_2)}.
\end{equation}

\begin{figure}
 \begin{center}
 \includegraphics[width=0.49\textwidth]{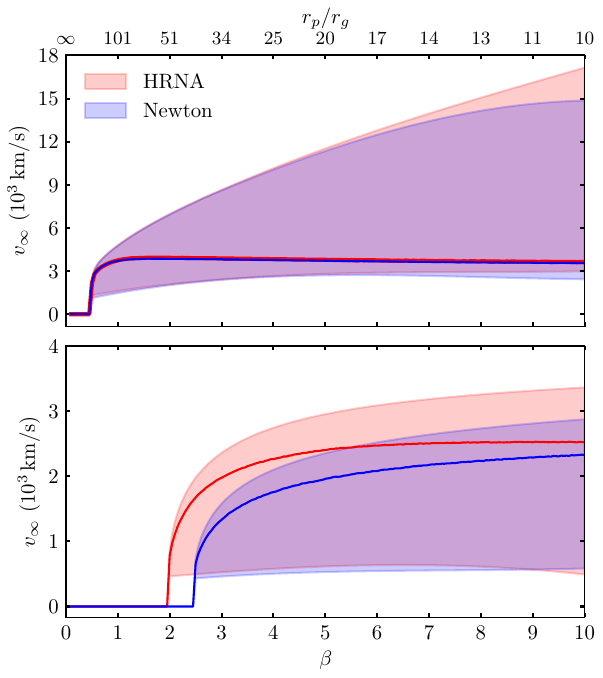}
 \end{center}
\caption{Asymptotic velocity $v_\infty$ of the ejected star (an HVS candidate) as a function of the impact parameter for prograde orbits (top panel) and retrograde ones (bottom panel). The shaded area shows the complete spread of the corresponding $v_\infty$ distribution, while average values are shown with solid lines in each case.}
\label{vel}
 \end{figure}

In Fig.~\ref{vel}, we present the resulting values of $v_\infty$ from the same Newtonian and HRNA encounters shown in Fig.~\ref{split}. The top panel illustrates the results of prograde encounters, while the bottom panel displays those for retrograde encounters. In both panels, shaded areas encompass all separation encounters, and solid lines represent the average values for each case. We have omitted 1PN results from this figure to maintain clarity, as the corresponding curves closely mirror those of the HRNA simulations.  

From Fig.~\ref{vel}, we observe a trend consistent with the previously discussed separation fraction: relativistic effects are notably more pronounced in retrograde encounters compared to prograde ones. However, in deeper prograde encounters ($\beta > 9$), the maximum velocity of the ejected star can be up to 2,000 km/s higher in HRNA simulations compared to their Newtonian counterparts.  Despite this significant difference, such high-velocity stars are rare and therefore have a minimal impact on the overall average velocity. 

Furthermore, in line with previous works (e.g.~Ref.~\cite{sari10}), we find that prograde encounters consistently result in higher asymptotic velocities compared to retrograde encounters. Additionally, in retrograde trajectories, relativistic encounters exhibit average velocities that are up to 30$\%$ higher than those observed in Newtonian simulations. 

\begin{figure}
 \begin{center}
  \includegraphics[width=0.49\textwidth]{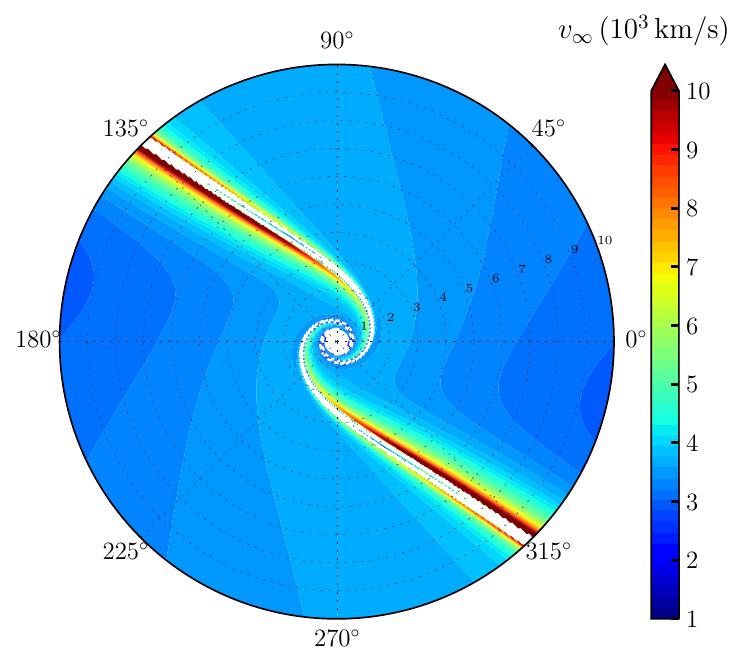}
 \end{center}
\caption{Polar representation of the distribution of $v_\infty$ as a function of $\beta$ and $\varphi$ for prograde HRNA encounters. Newtonian and 1PN results exhibit qualitative similarity. The white empty belt corresponds to the fraction of surviving binaries. Labels along the radial axis correspond to the impact parameter $\beta$ while angular labels correspond to the initial phase $\varphi$. }
\label{beta-phi-pro}
 \end{figure}

Fig.~\ref{beta-phi-pro} shows the distribution of $v_\infty$ in polar representation as a function of the two model parameters, $\beta$ and $\varphi$, for HRNA prograde encounters. Distributions from to Newtonian and 1PN simulations are essentially indistinguishable in this case. This reaffirms that relativistic effects are scarcely noticeable in such encounters. We can also notice that encounters with the largest values of $v_\infty$ are concentrated along a narrow strip within the white empty belt formed by surviving encounters. 

\begin{figure*}
 \begin{center}
  \includegraphics[width=0.49\textwidth]{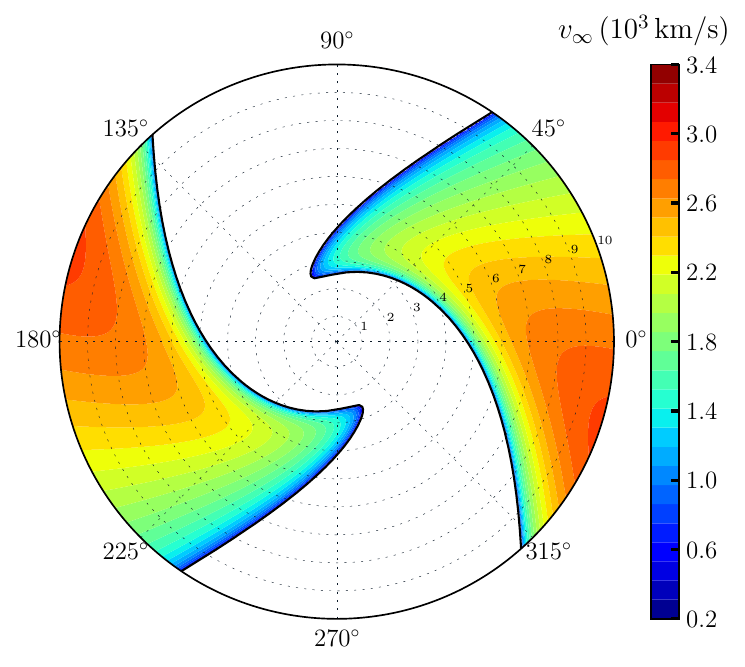}
  \includegraphics[width=0.49\textwidth]{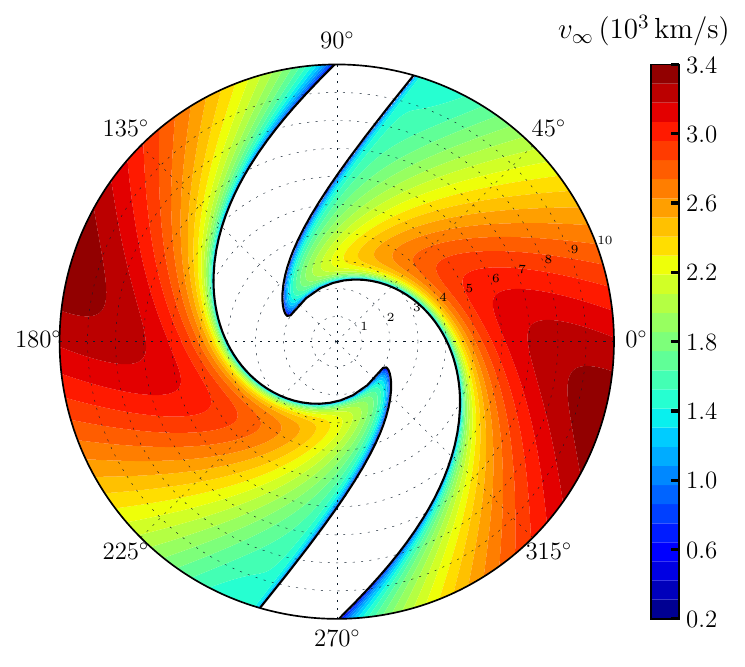}
 \end{center}
\caption{Same as Fig.~\ref{beta-phi-pro}, but now for retrograde trajectories for Newtonian (left-hand panel) and HRNA (right-hand panel) encounters. Results from 1PN simulations are very similar to HRNA ones. }
\label{beta-phi-retro}
 \end{figure*}

In Fig.~\ref{beta-phi-retro} we now show the distributions of $v_\infty$ in the case of retrograde trajectories for Newtonian and HRNA simulations. In contrast to prograde trajectories, surviving encounters in this case cluster together within a simply connected belt-like region (white empty area). In accordance with the results shown in Fig.~\ref{vel}, by comparing the left and right-hand panels, it is clear that relativistic encounters are more effective at separating binaries and, moreover, that they imprint larger asymptotic velocities on the ejected star.

\begin{figure*}
 \begin{center}
  \includegraphics[width=0.48\linewidth]{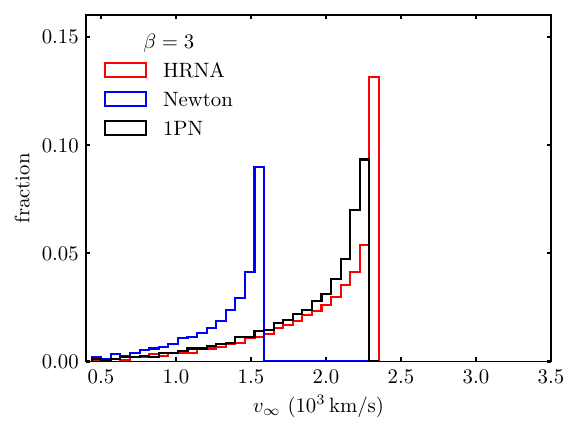}
  \includegraphics[width=0.48\linewidth]{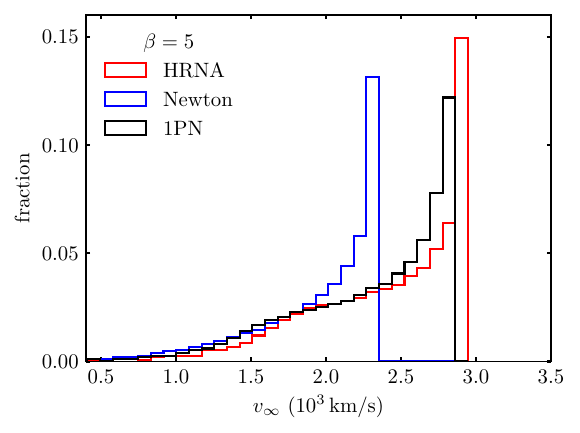}\\
  \includegraphics[width=0.48\linewidth]{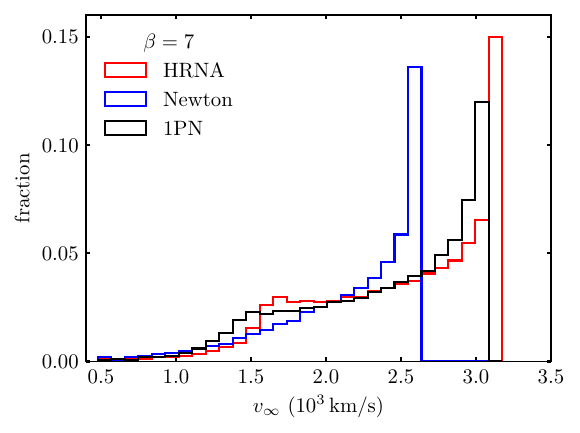}
  \includegraphics[width=0.48\linewidth]{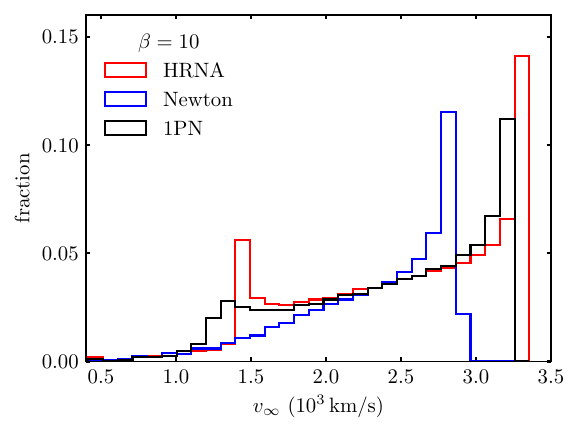}
 \end{center}
\caption{Comparison between the resulting distributions of $v_\infty$ for Newtonian, HRNA and 1PN retrograde encounters for different values of the impact parameter $\beta$.}
\label{vdist}
 \end{figure*}

In Fig.~\ref{vdist} we compare in more detail the distribution of $v_\infty$ of Newtonian, HRNA and 1PN simulations along retrograde trajectories, for values of $\beta = 3,\, 5,\, 7,\, 10$. In this figure we corroborate that relativistic encounters result in more and faster HVS candidates, with maximum velocities of up to $30\%$ larger than for Newtonian encounters. Moreover, we can also note that for $\beta\lesssim 5$, $v_\infty$ follows roughly a power law distribution with qualitatively similar shapes for Newtonian and relativistic encounters. Instead, for $\beta> 5$ a bimodal distribution starts developing for relativistic encounters.

\subsection{Surviving binaries and close encounters}

As the binary system approaches the central BH, tidal forces perturb the binary orbit by modifying both its eccentricity and the separation between the stars. These effects become larger and more dominant as $\rc \rightarrow\rt$. Even binary systems that survive the encounter can end up with their orbital elements highly modified.

\begin{figure*}
 \begin{center}
  \includegraphics[width=0.49\textwidth]{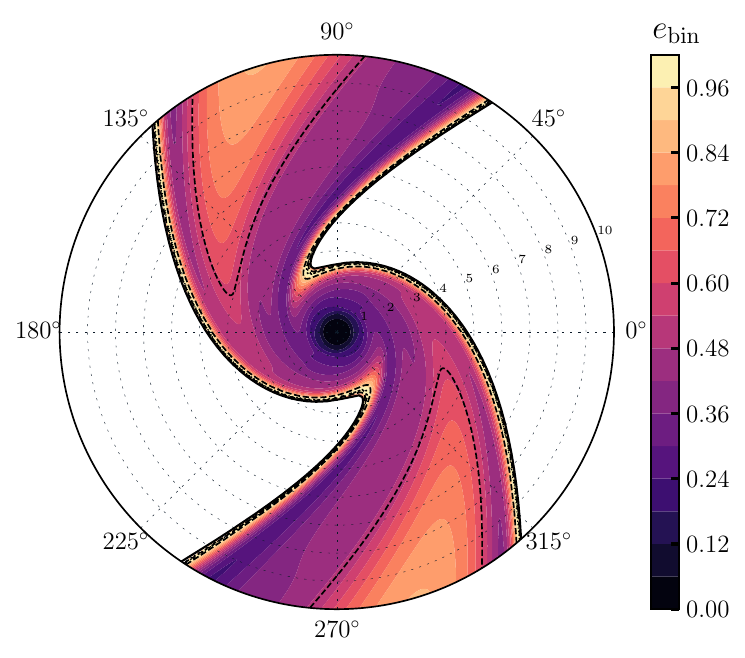}
  \includegraphics[width=0.49\textwidth]{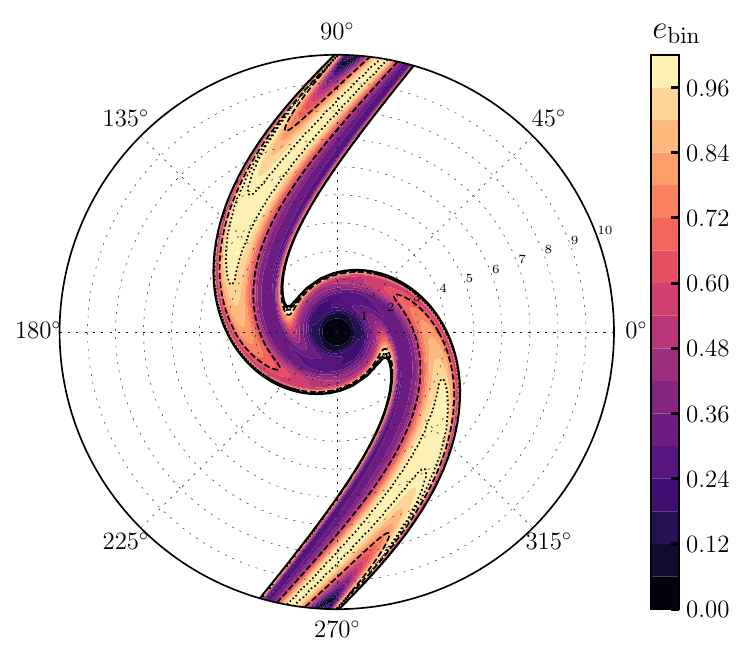}
 \end{center}
\caption{Binary eccentricity of the surviving encounters as a function of $\beta$ and $\varphi$ for retrograde trajectories. The left-hand panel shows Newtonian encounters and the right-hand panel HRNA ones.}
\label{ecc-polar}
 \end{figure*}

In Fig.~\ref{ecc-polar} we show the eccentricity of surviving binaries $\eb$ as a function of the model parameters $\beta$ and $\varphi$. Results of Newton and HRNA retrograde simulations are shown side-by-side. The distribution obtained for 1PN simulations is qualitatively very similar to the HRNA results. From this figure we see that relativistic simulations result in larger values of the final binary eccentricity. 

\begin{figure*}
 \begin{center}
  \includegraphics[width=0.49\textwidth]{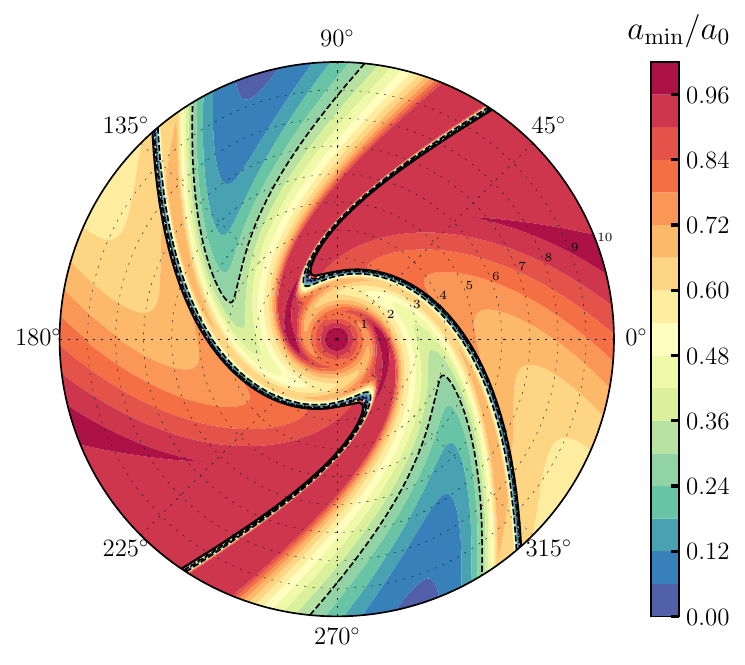}
  \includegraphics[width=0.49\textwidth]{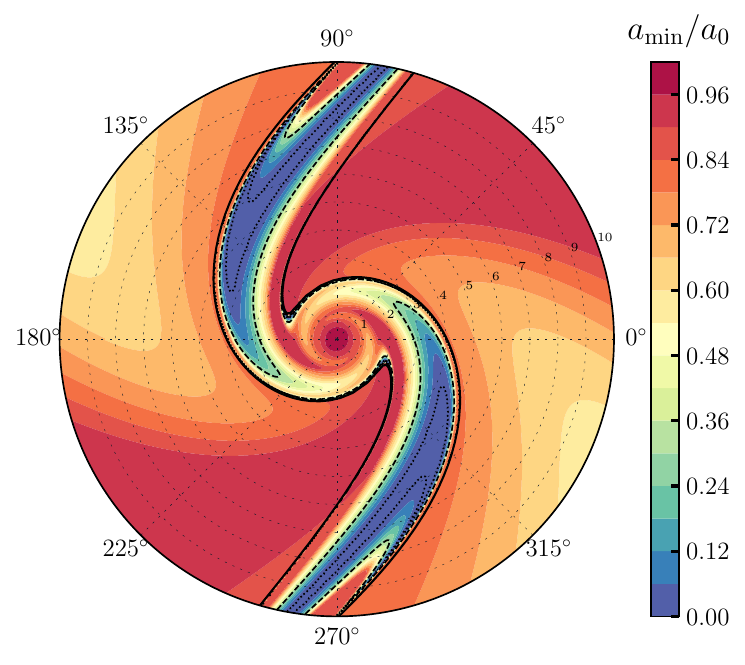}
 \end{center}
\caption{Minimum binary separation attained during the evolution as a function of $\beta$ and $\varphi$ for retrograde trajectories. The left-panel shows Newtonian encounters while the right-panel shows HRNA ones. The belt delimited by two solid lines on each plot corresponds to the surviving binaries. Encounters inside the dashed contour line ($a<0.3\,a_0$) would result in collisions between two main sequence stars (type 2 encounter), while those inside the dotted line ($a<0.01\,a_0$) would result in collisions between two white dwarfs (type 1 encounter).  }
\label{minsep-retro}
 \end{figure*}

In Fig.~\ref{minsep-retro} we show the minimum binary separation attained during the evolution of retrograde trajectories. By comparing Figs.~\ref{ecc-polar} and \ref{minsep-retro}, we can see a tight correlation between $\eb$ and $a_{\min}$: those binaries that experience the largest changes in their eccentricities also face the closest encounters during their evolution. 

\begin{figure}
 \begin{center}
\includegraphics[width=0.49\textwidth]{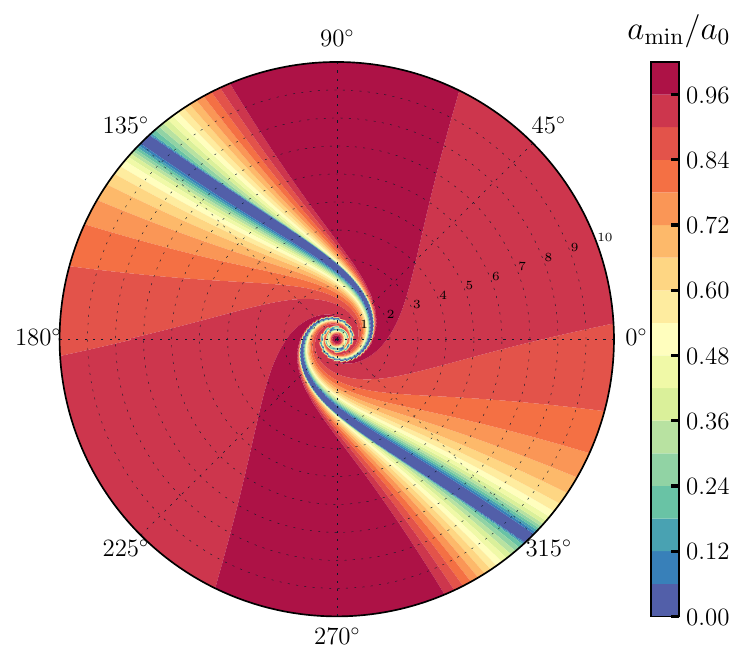}
 \end{center}
\caption{Minimum binary separation attained during the evolution as a function of $\beta$ and $\varphi$ for prograde HRNA encounters.}
\label{minsep-pro}
 \end{figure}

In Fig.~\ref{minsep-pro} we show the resulting distribution of $a_{\min}$ for HRNA prograde encounters. The corresponding results obtained for Newtonian and 1PN simulations are qualitatively very similar. From Figs.~\ref{minsep-retro} and \ref{minsep-pro},   it is apparent that close encounters are predominantly found among surviving encounters. 

\begin{figure}
 \begin{center}
  \includegraphics[width=0.49\textwidth]{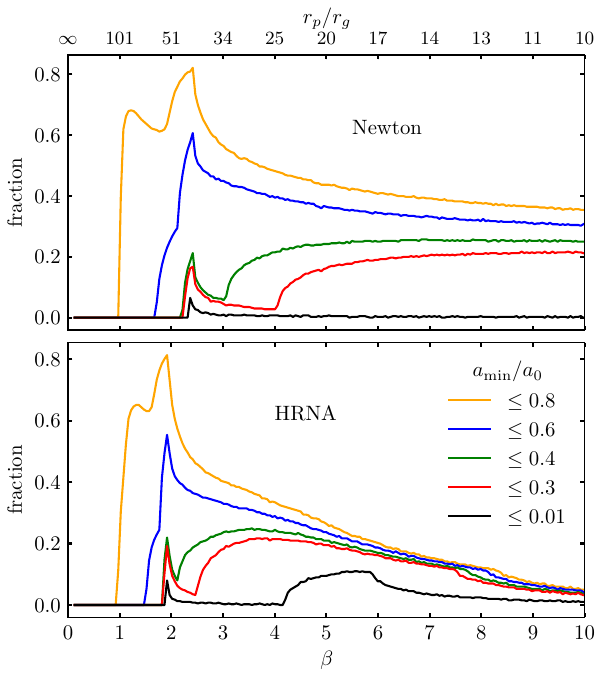}
  \caption{Cumulative fraction of binaries that reach a certain minimum separation as a function of the impact parameter $\beta$. The top-panel shows Newtonian encounters and the bottom-panel shows relativistic ones. In both cases, only results for retrograde encounters are shown.}
  \label{frac-amin}
 \end{center}
\end{figure}

In Fig.~\ref{frac-amin} we show the cumulative fraction of binaries that reach a minimum separation \mbox{$a_{\min}/a_0 \leq 0.8,\,0.6,\,0.4,\,0.3,\,0.01$}. By comparing Newtonian and HRNA results, we confirm the previous observation that relativistic simulations produce a larger fraction of close encounters. From this figure, we observe that the largest fraction of close encounters occurs for values of $\beta$ that coincide with the onset of binary separation, i.e.~for $\beta\simeq2,\,2.5$ for HRNA and Newtonian simulations, respectively. Moreover, we can also notice a prominent overabundance of very close encounters with $a_{\min}\leq 0.01\,a_0$ for HRNA encounters in the window $4.2\le\beta\le7$.

\begin{figure}
 \begin{center}
  \includegraphics[width=0.49\textwidth]{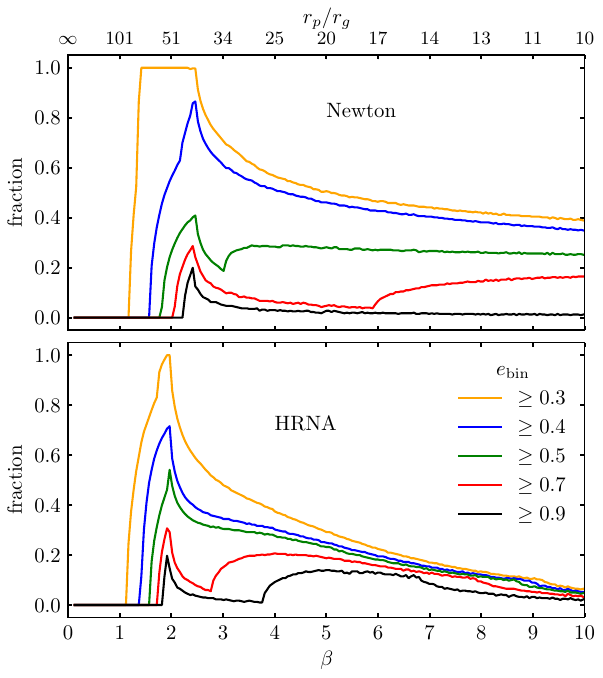}
  \caption{Cumulative fraction of binaries with a final eccentricity $e_{\mathrm{bin}}$ greater than five reference values as a function of the impact parameter $\beta$. The top-panel shows Newtonian encounters and the bottom-panel shows relativistic ones. In both cases, only results for retrograde encounters are shown.}
  \label{frac-ebin-1}
 \end{center}
\end{figure}

As a related quantity of interest, in Fig.~\ref{frac-ebin-1} we show the cumulative fraction of surviving binaries that end up with an eccentricity \mbox{$\eb \geq 0.3,\,0.4,\,0.5,\,0.7,\,0.9$}. From this figure we see that relativistic encounters can produce up to 15$\%$ of binaries with $\eb\geq 0.9$.

\subsection{Stellar collisions}

Once an astrophysical scenario is selected (e.g.,~one of the three type of encounters listed at the end of Section~\ref{S3}), the physical size of each binary member becomes a relevant parameter. Consequently, a fraction of the closest encounters discussed above could potentially lead to direct collisions between the binary members. Since this study treats each star as a point particle, detailed tracking of such collisions is unfeasible. However, given a stellar model, we can identify the fraction of potential collisions.  

Let now consider the example of a type 1 encounter (a 10$^6\Msun$ BH and a binary of $0.5\Msun$ white dwarfs, each with a radius $R_*\simeq 0.015 R_\odot \simeq  0.007 \,\rg$ \cite{ST83}). In this case, with a given initial separation of $a_0 = 0.01\,\mathrm{au} = 1.013\, \rg$, we can expect collisions whenever $a_{\min}\leq 0.01\,a_0$. Encounters that attain this or smaller values of $a_{\min}$  are shown enclosed within a dotted contour line in Fig.~\ref{minsep-retro}. This reference value is also shown as a cumulative fraction in Fig.~\ref{frac-amin}. From these figures we notice that this type of collisions are rare, especially for Newtonian encounters. As mentioned previously, relativistic simulations seem to predict a narrow window of the impact parameter $4.2\lesssim\beta\lesssim 7$ where these encounters are more abundant, attaining up to a $10\%$ of the total fraction. 

On the other hand, for a type 2 encounter (a 10$^7\Msun$ BH and a binary formed by two main sequence stars, each with a mass of 5$\Msun$ and a radius of $R_*\simeq 3R_\odot \simeq  0.14 \,\rg$ \cite{KWW}), with an initial separation of $a_0 = 0.1\,\mathrm{au} = 1.013\, \rg$, we should expect a collision whenever $a_{\min}\leq 0.3\,a_0$. This reference value of $a_{\min}$ is shown in Fig.~\ref{minsep-retro} by a dashed-line black contour, as well as in  Fig.~\ref{frac-amin}. From these figures we see that collisions between main sequence stars can be found for fraction of up to $20\%$ of encounters among both Newtonian and HRNA simulations. Note, however, that in the former case these collisions take place predominantly from $\beta=4$ onward while, in the latter, they occur for $2.4\lesssim\beta\lesssim 7.5$. 

It is worth noting at this point that type 2 encounters with $\beta\geq 5$ are also prone to producing double tidal disruption events, since in this case the tidal radius for individual encounters is $\rt \simeq 18\,\rg$ \cite{tejeda17}. Double tidal disruption events have been studied by e.g., Refs.~\cite{mandel15,bonnerot19}.  

Finally, for a type 3 encounter involving a 10$^8\Msun$ BH and a binary formed by two 50$\Msun$ BHs, collisions would occur only for an extremely small fraction of encounters with $a_{\min} \leq 10^{-6}$. We thus conclude that a type 3 encounter is highly unlikely to result on a BH collision. Nonetheless, as depicted in Fig.~\ref{frac-ebin-1}, such an encounter can lead to the formation of binaries with large eccentricities $\eb>0.9$. We discuss the future evolution of these eccentric, compact binaries in the next sub-section.

\subsection{Binary mergers}

A compact binary, composed of two 50$\Msun$ BH in circular orbit around each other with an initial separation of $a_0 = 1$\,au, is expected to merge due to gravitational wave emission within a Peters lifetime \cite{peters64} of
\begin{equation}
\begin{split}
    T_\mathrm{gw,0} & =  \frac{5\,a_0^4}{64\, m_b^3} \\ 
    &\simeq 1.3 \times 10^{12} 
    \left(\frac{a_0}{1\, \mathrm{au}}\right)^4
    \left(\frac{\mb}{100\Msun}\right)^{-3}\,\mathrm{yr}.
\end{split}
\end{equation}
This timescale being two orders of magnitude greater than the Hubble time, justifies our initial treatment of the binary within the Newtonian regime. 

However, should the binary survive a close encounter with the SMBH, its modified orbital parameters ($a_b$, $e_b$) may push it into a relativistic regime \cite{antonini12}. This shift is characterized by a substantially reduced Peters lifetime $ T_\mathrm{gw}$, potentially turning the binary into a gravitational wave source detectable by instruments like LIGO \cite{ligo16}.  

\begin{figure}
 \begin{center}
  \includegraphics[width=0.49\textwidth]{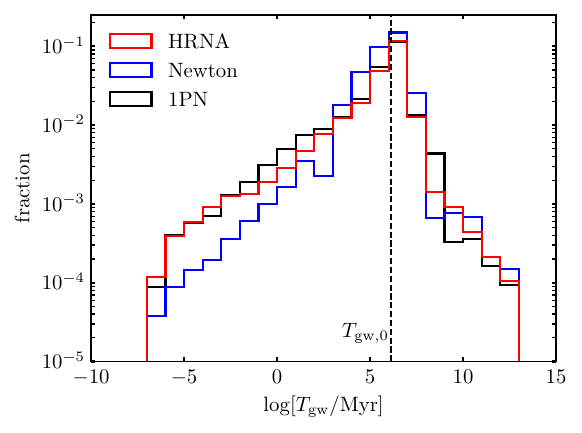}
  \caption{Comparison of the distributions of coalescence times $T_\mathrm{gw}$ for surviving binaries obtained from three different gravitational models. The distributions are expressed as a fraction of the total number of encounters. }
  \label{hist_Twg}
 \end{center}
\end{figure}

In Fig.~\ref{hist_Twg}, we present a comparison of the post-encounter distribution of $T_\mathrm{gw}$ for the three gravitational models. The fraction of each outcome is calculated relative to the total number of encounters. The time $T_\mathrm{gw}$ has been determined using Eq.~(5.14) from \cite{peters64}, which is applicable to orbits with arbitrary eccentricities $0<e_b<1$.  

The results clearly demonstrate that HRNA simulations predict a higher fraction of surviving binaries with relatively short coalescence times compared to Newtonian simulations. Specifically, HRNA predicts twice as many surviving binaries with $T_\mathrm{gw} \leq 10^8\,$yr (1.4$\%$ versus 0.76$\%$), and almost four times as many for $T_\mathrm{gw} \leq 10^4\,$yr (0.33$\%$ versus 0.08$\%$). 

Furthermore, if a binary BH remains bound to the SMBH following the tidal encounter, it can become an interesting dual source of gravitational waves: with high-frequency emissions associated with the binary's internal motion detectable by LIGO, and low-frequency emissions from the binary as whole acting as an EMRI source, potentially detectable by LISA \cite{addision19,chen19,AS23}. 

\subsection{Subsequent encounters}

As demonstrated by the previous results, even binaries that survive an encounter with the central BH can undergo significant alterations to both their internal and CM orbital elements. Notably, a considerable portion of surviving encounters culminates in binaries becoming gravitationally bound to the BH ($\Ec <0$). Consequently, these binaries are bound to return to the vicinity of the central BH for at least a second encounter, during which the system could potentially become tidally separated \cite{antonini10}. 

\begin{figure}
 \begin{center}
  \includegraphics[width=0.49\textwidth]{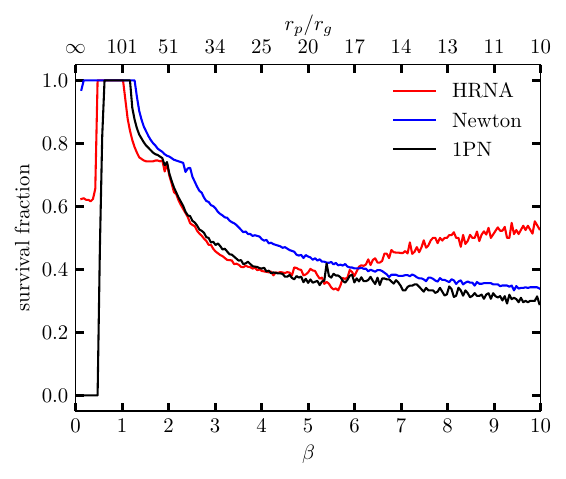}
  \caption{Fraction of surviving encounters that end up bound to the central BH after a first passage as a function of $\beta$. As a result, these binaries are bound to return to the vicinity of the central BH for a second encounter. Only results for retrograde encounters are shown for the three gravity laws considered in this work.} 
  \label{surv-fraction}
 \end{center}
\end{figure}

In Fig.~\ref{surv-fraction} we show the relative fraction of binaries among surviving encounters that remain bound to the BH as a function of the impact parameter for retrograde orbits and for each gravity law. 

One notable observation from Fig.~\ref{surv-fraction} is the substantial disparities between Newtonian and relativistic simulations, consistent with earlier findings. Conversely, results from 1PN and HRNA simulations are in quite good agreement with each other for $\beta\leq 6$.  Additionally, it is apparent that for $\beta>8$, over $50\%$  of HRNA surviving encounters are bound to return, contrasting with approximately $30\%$ of Newtonian surviving encounters exhibiting the same behavior. 

Another interesting result is that despite all retrograde encounters with $0.5\lesssim\beta\lesssim 1.2$ surviving (c.f.~Fig.~\ref{split}), essentially all of them become bound to the central BH after their initial encounter. Consequently, they are compelled to undergo a second close encounter. Notably, this observation remains valid for approximately  $70\%$ of the binaries prior to the onset of actual separations during the first encounter  ($\beta\simeq 2,\,2.5$ for relativistic and Newtonian encounters, respectively). 

The extensive computational time required to numerically evolve any of these surviving binaries bound to return for a second encounter with the central BH lies beyond the scope of the current study. We leave a detailed study of these cases to future research. 

\section{Summary and conclusions}
\label{S5}

We have studied the role of general relativistic effects on the tidal separation of binary stellar systems by non-rotating supermassive BHs (as described in Schwarzschild spacetime). These effects have been modeled using the Hybrid Relativistic Newtonian Approximation (HRNA), first introduced in Ref.~\cite{tejeda17}. Within this approximation, a Newtonian description of the binary interaction is combined with an exact relativistic treatment of the acceleration exerted by the central BH on each binary member. 

In order to perform a first exploration of  this type of tidal interactions with the HRNA method, we have adopted a restricted parameter space. Specifically, we have considered only equal mass binary members on an initial circular orbit, approaching the central BH along a parabolic-like trajectory. All motions were restricted to the equatorial plane of the central BH. Additionally, we have considered encounters with a fixed mass ratio of \mbox{$\mb/M = 10^{-6}$} and an initial binary separation of $a_0 = 1.013\,\rg$. 

We have systematically explored the resulting three dimensional parameter space spanned by the impact parameter $\beta\in[1,\,10]$, initial orbital phase $\varphi \in [0,2\pi]$, and binary sense of rotation (i.e.~prograde and retrograde orbits). 

To understand the effects of general relativity in these encounters, as well as to gauge the adequacy and limitations of the HRNA method, we have compared in detail the outcome of simulations performed with this method against the results of Newtonian and first order, Post-Newtonian (1PN) simulations for the same set of encounter parameters.  

Relativistic effects are generally subdominant \mbox{($<0.1\%$)} for prograde encounters in quantities such as separation fraction, post-encounter eccentricity distribution, average velocity of the ejected star, among others. In these cases, results from Newtonian, 1PN and HRNA simulations show good agreement.  

The exception occurs in very deep encounters ($\beta \geq 9$), where we observe a few cases where ejected stars reach maximum asymptotic velocities up to 2,000 km/s greater than those predicted by Newtonian simulations (see Fig.~\ref{vel}).

On the other hand, relativistic effects become crucial for retrograde encounters, where HRNA simulations notably diverge from Newtonian results. Results from both HRNA and 1PN simulations  exhibit strong agreement for impact parameters up
to $\beta = 5$. However,  for deeper encounters ($5\leq\beta\le10$), deviations between the two methodologies increase, reaching up to $10\%$.  

In retrograde encounters, tidal separation starts at lower values of the impact parameter ($\beta\simeq 2$) in both HRNA and 1PN simulations, in contrast to Newtonian simulations ($\beta\simeq 2.5$). Additionally, for a given $\beta$, relativistic simulations lead to separation fractions up to $30\%$ larger than those derived from Newtonian simulations (see Fig.~\ref{split}). 

Similarly, ejected stars during relativistic simulations of retrograde encounters attain average asymptotic velocities up to $500\,$km/s higher than those ejected during Newtonian simulations (see Figs.~\ref{beta-phi-retro} and \ref{vdist}). 

The observation that relativistic effects are more pronounced in retrograde encounters compared to prograde ones can be understood through their distinct dynamics. In prograde trajectories, tidal separation typically occurs abruptly and close to pericenter passage, as illustrated in Fig.~\ref{sep-P}. In contrast, during  retrograde encounters, the tidal field initially acts to slow down and reverse the binary's sense of rotation, as shown in Fig.~\ref{sep-R}. This process not only delays the eventual separation but also prolongs the binary's interaction with the BH's tidal field, thus amplifying the overall impact of these encounters. 

Binaries that survive the tidal interaction with the SMBH do not emerge unaltered. We observe that the majority of these binaries are left bound to the central BH after the tidal interaction (see Fig.~\ref{surv-fraction}). This implies that they are likely to undergo one or more subsequent close encounters with the SMBH in the future, potentially leading to their eventual separation. 

This significant observation might go unnoticed in studies based on restricted three-body simulations \citep[e.g.~][]{kobayashi12,bkrs18}, which assume that the CM motion remains unaffected during the tidal interaction. Our findings underscore the importance of considering changes to the CM motion in simulations to fully capture the dynamics and long-term evolution of binary systems in the vicinity of SMBHs.

Moreover, when considering the finite radii of the binary members, a significant fraction of these encounters that apparently survive could, in fact, result in stellar collisions (see Fig.~\ref{frac-amin}). For the explored parameters, both Newtonian and relativistic simulations indicate the potential for collisions between two main sequence stars. However, encounters resulting in collisions between two white dwarfs are predominantly found among relativistic encounters, especially for $2.4 \leq \beta \leq 7.5$.

Surviving binary systems composed of two BHs are significantly affected by their tidal encounters with a SMBH, driving them into a regime where gravitational wave emission becomes a critical factor in their future evolution. These interactions can drastically reduce the coalescence times of the binaries ($T_{\mathrm{gw}}$), with reductions up to 13 orders of magnitude shorter than those without such an interaction (see Fig.~\ref{hist_Twg}).
Moreover, HRNA simulations predict that the number of surviving binaries with $T_{\mathrm{gw}} \leq 10^8$\,yr is twice that predicted by Newtonian simulations.

In conclusion, we have conducted a first study of the influence of general relativity on tidal encounters between binary systems and SMBHs. We have demonstrated that relativistic effects are crucial for accurately predicting the dynamics of such encounters, particularly in retrograde orbits where relativistic effects are most pronounced.

Additionally, we have validated the HRNA method as a reliable and efficient tool for examining relativistic effects in astrophysical systems. Our findings provide valuable insights into the dynamics of binary tidal separations that extend across a range of phenomena. These include increased rates of binary separation, variations in the velocities of ejected stars, and heightened probabilities of stellar collisions and binary mergers; 
each essential for understanding stellar dynamics near SMBHs.

The implications of our research are relevant to studies on the properties and dynamics of stellar populations in the vicinity of SMBHs as well as for gravitational wave astronomy. Moreover, a deeper understanding of the dynamics of star systems around BHs can provide crucial insights into the evolution of galaxies and the intrinsic properties of BHs themselves. Furthermore, our results support the necessity for incorporating relativistic models in simulations of galactic centers, especially to predict the characteristics of hypervelocity stars.

For future research, it would be beneficial to expand this study to include non-equatorial orbits and to explore the effects of different binary mass ratios and orbital configurations. Additionally, investigating the influence of the BH's spin could provide deeper insights into the complex interplay between stellar dynamics and BH physics, potentially revealing new facets of these complex systems.

\section*{Acknowledgments}

We thank Olivier Sarbach for insightful discussions and critical comments on the manuscript. ET acknowledges James Guillochon and Stephan Rosswog for useful comments on an earlier version of the text. This work was partially supported by CONAHCYT Ciencia de Frontera Project No.~376127 ‘Sombras, lentes y ondas gravitatorias generadas por objetos compactos astrof\'isicos’.

\appendix

\section{Self-consistency criterion}
\label{Aalpha}

In \eq{alpha} the parameter $\alpha=a/\mathcal{R}$ was introduced as a self-consistency criterion for the applicability of the HRNA method. When $\alpha\ll 1$, the length scale on which the binary's self-gravity operates ($a$) is much smaller than the local radius of curvature of the spacetime ($\mathcal{R}$) and, thus, the assumptions behind the HRNA method are expected to be valid.

\begin{figure}
    \centering
    \includegraphics[width=1\linewidth]{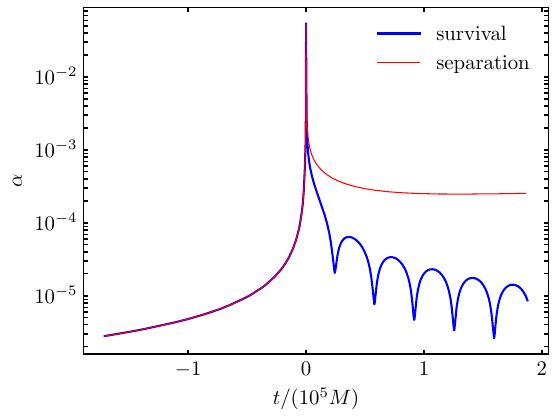}
    \caption{
    Instantaneous $\alpha$ value for two representative encounters of separation and survival of the binary system. The simulations correspond to relativistic encounters (evolved under the HRNA method) in retrograde orbits with impact parameter $\beta = 8.5$, initial phase $\varphi = 1.7$ (binary survives) and $\varphi = 1.8$ (binary separated). Time zero corresponds to the maximum of the tidal interaction, i.e.~the CM pericenter passage.}
    \label{fig:alpha}
\end{figure}

In Fig.~\ref{fig:alpha} we show representative examples of the time evolution of $\alpha$ for two encounters with $\beta = 8.5$, one resulting in binary separation and the other in survival. From this figure it is clear that $\alpha\sim10^{-5}$ for most of the simulation time, even though values as large as $\alpha\sim 0.1$ are reached for a brief interval of time around pericenter.

\begin{figure}
    \centering
    \includegraphics[width=1\linewidth]{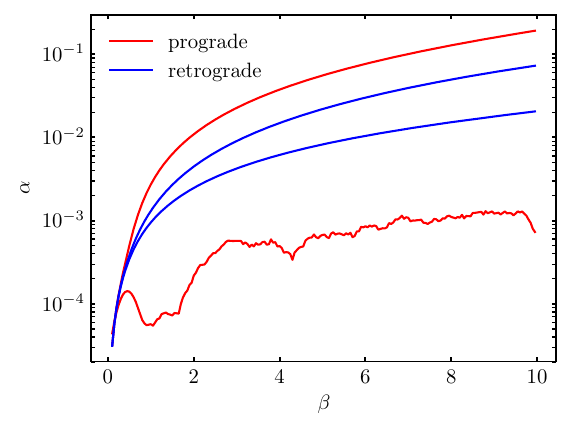}
    \caption{Maximum and minimum values of the parameter $\alpha=a/\mathcal{R}$ achieved during pericenter passage among the 400 simulations run for each value of the impact parameter $\beta$. Results for prograde (retrograde) trajectories are shown in red (blue) color lines.}
    \label{fig:alphabeta}
\end{figure}

In Fig.~\ref{fig:alphabeta} we show the maximum and minimum values of $\alpha$ achieved at pericenter among the 400 phases explored for each value of the impact parameter. From this figure we observe that there is a larger spread of values for prograde encounters as compared to retrograde ones. 

From these results we can conclude that the condition in \eq{alpha} is satisfied for most of the simulation time of each encounter. It remains unclear, however, whether the brief interval of time during which $\alpha\sim 0.1$ could compromise the consistency of the method.

\section{Relativistic time dilation in HRNA}
\label{AH}

In this appendix, we showcase the ability of the HRNA method  to accurately capture the effects arising from both special and general relativistic dilation of time. 

To more effectively isolate this phenomenon, we take a harmonic oscillator as the self-interacting system in the evolution equation \eq{CoordtEqs}.
That is, we consider two equal masses $m_1 = m_2 = m$ coupled by a Hooke-type potential given by
\begin{equation}
    \Phi =\frac{1}{2}k\left ( \left| \bm{x}_{2}-\bm{x}_{1} \right|-l_{0} \right )^{2},
\end{equation}
where $l_0$ is the equilibrium distance between the two masses. 

In order to eliminate confounding effects from torques and rotational motion, the two masses are aligned perpendicularly to the radial direction. Additionally, we take as fixed parameters $k/m = 1/5$ and $l_0=3\,r_g/40$. In all simulations reported bellow, the oscillator is initially perturbed with an elongation of $\left| \bm{x}_{2}-\bm{x}_{1} \right|=4\,l_{0}/3 $.

We first focus on the  gravitational redshift component of the dilation by fixing the center of mass of the oscillator at a distance $r$ from the central BH (as if held in place by a rocket exactly counteracting the gravitational attraction of the BH). 
The natural frequency of the oscillator, as measured by a local co-moving observer, is given by 
\begin{equation}
  f_{0}=\frac{1}{2\pi }\sqrt{\frac{2k}{m}}.
\end{equation}
Nevertheless, during a numerical simulation of this system using the HRNA method, the obtained frequency measurement should agree with the description done by a distant observer using coordinate time $t$. In other words, in this scenario, we should expect to measure a gravitationally redshifted frequency given by
\begin{gather}
  f_\mathrm{st} = f_0/\Gamma_\mathrm{st}, 
\label{redshift}
  \\  \Gamma_\mathrm{st} = \left( 1 - \frac{2M}{r}   \right)^{-1/2},
\end{gather}
where $\Gamma_\mathrm{st}$  is the Lorentz factor for a static observer at a fixed $r$ (c.f.~Eq.~\ref{gamma}).

\begin{figure}[h!]
\begin{center}
\includegraphics[width=0.98\linewidth]{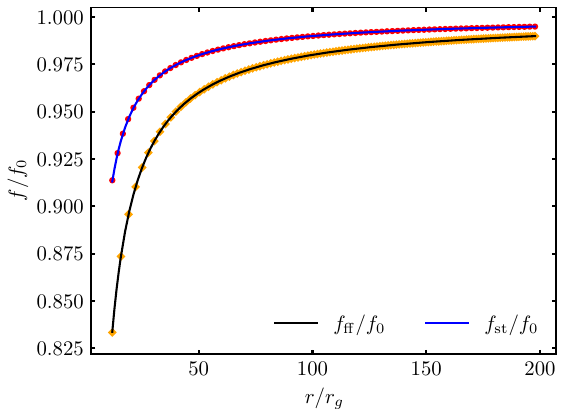}
    \caption{    Relativistic time dilation of a harmonic oscillator evolving under the HRNA method. 
    A first series of experiments with the CM of the oscillator fixed at a given radius is shown with red circles. The result of a second experiment with the oscillator in radial free-fall towards the central BH is shown with orange diamonds.
    For comparison, the analytically expected trends in each case are indicated by  continuous blue and black lines, respectively.\label{fig:bamb}}
\end{center}
\end{figure}

In Fig.~\ref{fig:bamb} we show the results of several realizations of this configuration for different radii between $r=10\,\rg$ and $r=200\,\rg$. The corresponding frequency at each point is computed as the inverse of the time lapse between successive maxima of the separation of the two masses.
As we can see from this figure, there is an excellent agreement between numerical measurements and the analytic expression in \eq{redshift}.

In a second numerical experiment combining both special and general relativistic effects, we take the oscillator in radial free-fall towards the central BH. Considering that the oscillator's CM starts from rest at infinity, in this case we expect to measure a frequency shift of
\begin{gather}
f_\mathrm{ff} = f_0/\Gamma_\mathrm{ff}, 
\label{ff}\\    \Gamma_\mathrm{ff} = \left( 1 - \frac{2M}{r}   \right)^{-1}.
\end{gather}

The obtained numerical results are also shown in Fig.~\ref{fig:bamb}, where we find again an excellent agreement between the numerically measured averaged frequency and \eq{ff}.

\section{Evolution equations in \\post-Newtonian gravity}
\label{AA}

The evolution equations for a general $N$-body system to first order in post-Newtonian approximation are given by \cite{Straumann2013,Merritt2016} 
\begin{widetext}
\begin{align}
    \ddot{\bm{x}}_a =& -\sum_{b \neq a} m_b \left( \dfrac{\bm{x}_{ab}}{r_{ab}^3} \right) \Bigg[  1-4 \sum_{c \neq a} \dfrac{m_c}{r_{ac}}  
    - 5 \dfrac{m_a}{r_{ab}} + \dot{\bm{x}}_a^2 - 4\,\dot{\bm{x}}_a \cdot \dot{\bm{x}}_b
      +\sum_{c \neq a,b} m_c \left(- \dfrac{1}{r_{bc}} + \dfrac{\bm{x}_{ab} \cdot \bm{x}_{bc} }{2\,r_{bc}^3} \right)  + 2\, \dot{\bm{x}}_b^2 - \dfrac{3}{2} \left( \dfrac{ \dot{\bm{x}}_b \cdot \bm{x}_{ab} }{r_{ab}} \right)^2 \Bigg]  \nonumber\\ 
    &~ - \dfrac{7}{2} \sum_{b \neq a} \left( \dfrac{m_b}{r_{ab}} \right) \sum_{c \neq a,b} \dfrac{m_c \bm{x}_{bc}}{r_{bc}^3} + \sum_{b \neq a} \dfrac{m_b}{r_{ab}^3} \bm{x}_{ab} \cdot \left( 4\,\dot{\bm{x}}_a - 3\, \dot{\bm{x}}_b \right) \left( \dot{\bm{x}}_a - \dot{\bm{x}}_b \right),
    \label{eab.1}
\end{align}
where ($a,\,b,\,c$) are particle labels and summations run over the whole $N$-body system.

In the special case of a restricted three-body system, in which one body is much more massive than the other two ($M\gg m_1,\,m_2$), we can consider the former as fixed at origin of coordinates. With this approximation, \eq{eab.1} reduces to
\begin{align}
    \ddot{\bm{x}}_{1} =&~ -m_{2} \dfrac{\bm{x}_{1 2}}{r_{1 2}^3} \Bigg[ 1 - \dfrac{4\,m_2}{r_{1 2}} - \dfrac{4 M}{r_{1}} + M \left(- \dfrac{1}{r_{2}} + \dfrac{\bm{x}_{1 2} \cdot \bm{x}_{2} }{2\,r_{2}^3} \right)  - \dfrac{5\, m_{1}}{r_{1 2}} + \dot{\bm{x}}_1^2 - 4\,\dot{\bm{x}}_1 \cdot \dot{\bm{x}}_2 + 2\,\dot{\bm{x}}_{2}^2 - \dfrac{3}{2} \left( \dfrac{ \dot{\bm{x}}_2 \cdot \bm{x}_{1 2} }{r_{1 2}} \right)^2 \Bigg]  \nonumber \\
    &~ -M \dfrac{\bm{x}_{1}}{r_{1}^3} \Bigg[ 1 - \dfrac{4\,m_2}{r_{1 2}} - \dfrac{4 M}{r_{1}} + m_{2} \left(- \dfrac{1}{r_{2}} - \dfrac{\bm{x}_{1} \cdot \bm{x}_{2} }{2\,r_{2}^3} \right) - \dfrac{5\, m_{1}}{r_{1}} + \dot{\bm{x}}_1^2  \Bigg] - \dfrac{7 M m_{2} \, \bm{x}_{2}}{2\, r_{2}^3}  \left( \dfrac{1}{ r_{1 2} } - \dfrac{1}{ r_{1} } \right) \nonumber \\
    &~ + \dfrac{m_{2}}{r_{1 2}^{3}} \bm{x}_{1 2} \cdot \left( 4\,\dot{\bm{x}}_1 - 3\,\dot{\bm{x}}_2 \right) \left( \dot{\bm{x}}_1 - \dot{\bm{x}}_2 \right) + \dfrac{4 M}{r_{1}^{3}} \left( \bm{x}_{1} \cdot  \dot{\bm{x}}_{1} \right) \dot{\bm{x}}_{1}, 
\end{align}
an equivalent equation of motion is obtained for $\ddot{\bm{x}}_{2}$ by interchanging the particle labels $1\leftrightarrow 2$.

\end{widetext}

\section{Initial conditions}
\label{AB}

In this Appendix we describe the initial conditions for the simulations presented in the body of the article.

For all three gravity laws considered in this work (Newton, HRNA, and 1PN), we prescribe the initial conditions for each member of the binary system according to the usual Newtonian expressions
\begin{gather}
x_{\{1,2\}} = x_{\rm cm} \pm \frac{m_{\{1,2\}}}{\mb}\,a\,\cos\varphi   
\label{eAB1},\\
y_{\{1,2\}} = y_{\rm cm} \pm \frac{m_{\{1,2\}}}{\mb}\,a\,\sin\varphi   ,\\
z_{\{1,2\}} = 0,\\
\dot{x}_{\{1,2\}} = \dot x_{\rm cm} \mp \frac{m_{\{1,2\}}}{\mb}\,a\,\dot \varphi\,\sin\varphi   ,\\
\dot{y}_{\{1,2\}} = \dot y_{\rm cm} \pm \frac{m_{\{1,2\}}}{\mb}\,a\,\dot \varphi\,\cos\varphi ,\\
\dot z_{\{1,2\}} = 0,
\end{gather}
where the upper (lower) sign corresponds to binary member 1 (2), and
\begin{equation}
   \dot \varphi = \pm\sqrt{\frac{\mb}{a^3}} \label{eAB7}
\end{equation}
is the Newtonian angular velocity for circular orbits, with the $\pm$ sign corresponding to either prograde or retrograde orbits, respectively. 
All quantities in \eqs{eAB1}-\eqref{eAB7} are to be evaluated at the initial time $t_i$ of the simulation.
Additionally, without loss of generality, we take $(x_{\rm cm},\,y_{\rm cm})=(r_0,0)$, as the initial position of the CM.

On the other hand, the initial CM velocity is different for each gravity law as we describe next. 

The resulting expressions are such that the resulting trajectory of a virtual test particle with mass $\mb$ placed at the binary's CM would follow a parabolic-like trajectory with a pericentre distance $\rp = \rt/\beta$ (cf.~Eq.~\ref{e1.0}) under the corresponding gravity law.

\subsection{Initial CM velocities in Newtonian gravity}

In Newtonian gravity, the motion of a test particle in orbit around a central mass $M$ is characterized by the existence of two conserved quantities: the specific energy $E$ and the specific angular momentum $L$. In the case of a parabolic orbit we have $E=0$ and $L = \sqrt{2M\rp}$ and the following initial CM velocities:
\begin{gather}
    \dot x_{\rm cm} = \dot r_0 = -\sqrt{\frac{2M}{r_0} - \frac{L^2}{r_0^2}},\\
    \dot y_{\rm cm} = \frac{L}{r_0}.
\end{gather}

\subsection{Initial CM velocities in\\Schwarzschild spacetime}

As discussed in more detail in the Appendix A of \cite{tejeda17}, the motion of a test particle in Schwarzschild spacetime in a parabolic-like trajectory has $E=0$ and $L=\sqrt{2M\rp^2/(\rp-2M)}$. In this case the initial CM velocities are given by
\begin{gather}
    \dot x_{\rm cm} = \dot r_0 = -\left(1-\frac{2M}{r_0}\right)\sqrt{\frac{2M}{r_0} - \frac{L^2}{r_0^2}\left(1-\frac{2M}{r_0}\right)},\\
    \dot y_{\rm cm} = \left(1-\frac{2M}{r_0}\right)\frac{L}{r_0}.    
\end{gather}

\subsection{Initial CM velocities in post-Newtonian gravity}

The evolution to first order in the post-Newtonian expansion of a test particle in a general trajectory around a gravitational mass $M$ is characterized by the conserved quantities:
\begin{gather}
    E = \frac{1}{2} v^2 - \frac{M}{r} + \frac{3}{8} v^4 + \frac{M}{2r} \left( 3\,v^2 +  \frac{M}{r} \right) , \label{EIH1}\\
    L = r^2\dot\phi \left( 1 + \frac{1}{2} v^2 + 3\frac{M}{r} \right), \label{EIH2}
\end{gather}
where $v = \sqrt{\dot r^2 + r^2\dot\phi^2}$.

In the case of a parabolic-like trajectory we have $E=0$. Taking this into account, and evaluating \eqs{EIH1} and \eqref{EIH2} at pericentre (where $\dot r = 0$), it follows
 \begin{gather}
     L = \rp\,v_\mathrm{p} \left( 1+ \frac{1}{2} v_\mathrm{p}^2 + 3 \frac{M}{\rp} \right),\\
     v_\mathrm{p}^2 = \frac{2(\rp + 3M )}{3\rp} \left(  \sqrt{1+\frac{3M(2\rp-M)}{(\rp + 3M )^2}} -1 \right).
 \end{gather}

 Next we evaluate \eqs{EIH1} and \eqref{EIH2} at $r_0$ and obtain
 \begin{gather}
  \dot x_{\rm cm} = - \sqrt{v_0^2 - \dot Y^2_0},\\
  \dot y_{\rm cm} = \frac{L}{r_0\left(1+v_0^2/2\right) + 3M},\\
 v_0^2 = \frac{2(r_0 + 3M )}{3r_0} \left(  \sqrt{1+\frac{3M(2r_0-M)}{(r_0 + 3M )^2}} -1 \right).
 \end{gather}

\section{Convergence with respect \\to the initial distance}
\label{AC}

As mentioned in Section~\ref{S3}, in this work we consider binary systems that approach the central BH along a parabolic-like trajectory. From a numerical perspective, we must ensure that our results are, as much as possible, independent from the choice of initial CM distance $r_0$.

To test the degree of  convergence of the separation fraction, the main focus of this work, with respect to the initial distance, we have run a series of simulations varying the initial distance $r_0$ by integral steps between $r_0=\rt$ and $r_0=100\,\rt$  for several values of the $\beta$ parameter and for the three gravity laws.

Following the same procedure described in Section~\ref{S3}, for each value of the initial distance $r_0$, we perform an initial exploration based on 400 simulations varying the initial orbital phase $\phi\in[0,\,\pi]$. We then employ the bisection algorithm to refine the search of the boundary between survival and separation encounters down to a precision of $10^{-4}$. The separation fraction is then defined according to \eq{sf}.

\begin{figure}
  \begin{center}
   \includegraphics[width=0.49\textwidth]{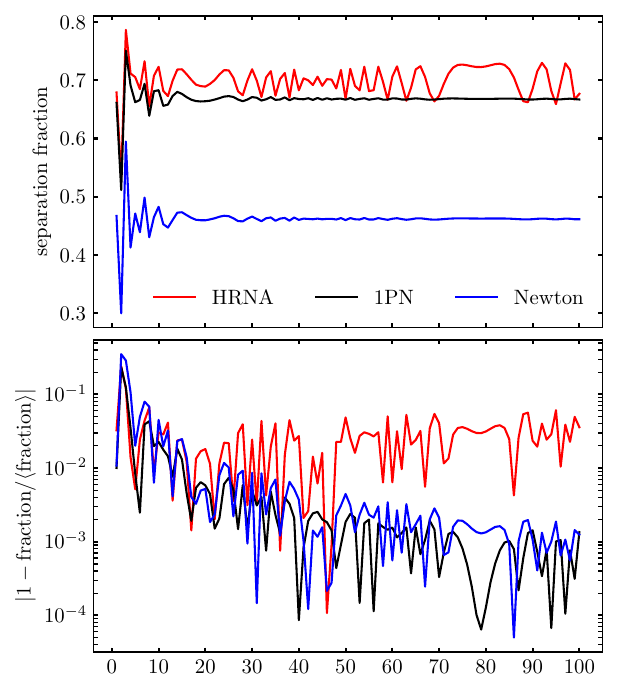}
  \end{center}
 \caption{ The top-panel shows the separation fraction as a function of the initial distance $r_0$ for an impact parameter $\beta = 4.7$. The bottom-panel shows the relative difference between the separation fraction and its global average. } 
 \label{fig:split-vs-r0}
  \end{figure}

In the top panel of Figure~\eqref{fig:split-vs-r0} we show the separation fraction obtained for the three gravity laws and for a representative value of $\beta=4.7$. The 
bottom panel shows the relative difference between the separation fraction at a given $r_0$ and its total average value.
Qualitatively similar results are obtained for other values of $\beta$.

In  Figure~\eqref{fig:split-vs-r0} we see a well-behaved convergence with respect to $r_0$ for both Newtonian and 1PN simulation. The relative difference  decreases as the ratio $r_0/\rt$ increases from 1 to 70. This trend stalls for $r_0 \gtrsim 70\,\rt$ as the obtained precision is comparable to the prescribed $10^{-4}$ accuracy of the bisection method.
On the other hand, there seems to be a weaker convergence for the HRNA results, as the average resulting precision stalls at around $10^{-2}$ already from $r_0 \gtrsim 20\,\rt$.

We have also explored the convergence of the simulations with respect to the final time of the simulation. 
Finishing the simulations at increasing final distances from the central BH, taking values of $r_f = 150\,\rt$ up to $r_f = 300\,\rt$, the obtained results do not show a strong dependence on this parameter.

The tests conducted in this work are not enough to clarify the reasons behind the weaker convergence of the HRNA method with respect to $r_0$ as compared to Newtonian and 1PN simulations. We speculate that this behavior might stem from the inherent approximated nature of HRNA. Studying the convergence properties of HRNA in more detail might help establishing its limits of applicability. 

\bibliography{references} 

\end{document}